\documentclass[lettersize,journal]{IEEEtran}
\usepackage{amsmath,amsfonts}
\usepackage{algorithmic}
\usepackage{algorithm}
\usepackage{array}
\usepackage{textcomp}
\usepackage{stfloats}
\usepackage{url}
\usepackage{verbatim}
\usepackage{graphicx}
\usepackage{cite}
\usepackage{booktabs}
\usepackage{subfigure}
\usepackage{subcaption}
\usepackage{xcolor}
\begin{document}

\title{Automated Discovery of Anomalous Features in Ultra-Large Planetary Remote
Sensing Datasets using Variational Autoencoders}

\author{Adam Lesnikowski$^a$, Valentin T. Bickel$^b$, Daniel Angerhausen$^{cd}$
\thanks{a Independent Researcher, Rhode Island, USA\\
        b Center for Space and Habitability, University of Bern, Bern, Switzerland\\
        c Institute for Particle Physics and Astrophysics, ETH Z\"urich, Z\"urich, Switzerland\\
        d Blue Marble Space Institute of Science, Seattle, Washington USA.\\
        Corresponding authors:\\
        valentin.bickel@unibe.ch, adam.lesnikowski@gmail.com}
}

\markboth{Journal of Selected Topics in Applied Earth Observation and Remote Sensing, forthcoming}
{Shell \MakeLowercase{\textit{et al.}}: A Sample Article Using IEEEtran.cls for IEEE Journals}

\maketitle

\begin{abstract}
The NASA Lunar Reconnaissance Orbiter (LRO) has returned petabytes of lunar high
spatial resolution surface imagery over the past decade, impractical for humans
to fully review manually. Here we develop an automated method using a deep
generative visual model that rapidly retrieves scientifically interesting
examples of LRO surface imagery representing the first planetary image anomaly
detector. We give quantitative experimental evidence that our method
preferentially retrieves anomalous samples such as notable geological features
and known human landing and spacecraft crash sites. Our method addresses a major
capability gap in planetary science and presents a novel way to unlock insights
hidden in ever-increasing remote sensing data archives, with numerous
applications to other science domains. We publish our code and data along with
this paper.
\end{abstract}


\section{Introduction}
\IEEEPARstart{W}{hat} do we do when scientific instruments generate vastly more
data than what is possible for humans to review? Here we seek to develop a
general method to retrieve scientifically interesting and strategically relevant
samples from very large remote sensing datasets in an automated way. Towards
this we work on the petabytes of image data collected by the NASA Lunar
Reconnaissance Orbiter (LRO) Narrow Angle Camera (NAC) over the past years, the
highest spatial resolution and quality image collection of the lunar surface
currently available, with more than two million images \cite{robinson2010}. Past
work on analyzing this dataset have heavily relied on manual review and
processing \cite{braden2014evidence,watters2019shallow,wagner2014}. Recently,
supervised learning techniques have systematically mapped geologically
interesting features such as fractured boulders and mass wasting locations on
regional and global scales \cite{ruesch2023,bickel2020impacts,bickel2022}. These
approaches have yielded important advances in our knowledge of lunar geological
processes, but face the bottleneck of available human labellers, while suffering
from a lack of generalization to samples which are interesting relative to the
dataset but unspecified in the manual labelled set, i.e., `anomalies'. As of
today, there exists no method that is able to identify anomalies in planetary
image data in an automated way. We wish to have some method to alleviate this
reliance on manual labels, and to do so in an agnostic way to rapidly find
anomalous samples unspecified at train time. Here by `anomalies' we mean samples
which are in a low-density part of our sample distribution, usually with high
scientific or strategic value, such as volcanic pits (skylights) and spacecraft
landing/crash sites. As one example, it took approximately two months of manual
review to find the Chandrayaan-2 crash site, after contact with the probe was
lost on September 7th 2019. Other crash sites, like the Chandrayaan-1 Moon
Impact Probe's impact site (2008), still remain to be found \cite{stooke2021}.

Image anomaly detection methods seek to find images which are anomalous with
regard to the bulk distribution in a given dataset or data stream.
\cite{giles2019,giles2020} demonstrated how traditional clustering algorithms
like DBSCAN (Density-Based Spatial Clustering of Applications with Noise) can be
used to recognize anomalies\,--\,such as unusual brightness fluctuations in
Boyajian's star (KIC 8462852)\,--\,in Kepler photometric light curve data. As an
alternative to traditional cluster algorithms, neural networks may be used to
learn what a typical image in some distribution looks like, in order to detect
atypical or anomalous images. Autoencoders and their extensions such as
variational autoencoders (VAEs) are neural network architectures that seek to
recreate their input as their outputs. \cite{kingma2013auto} presents VAEs, but
does not apply it to anomaly detection. In turn, \cite{an2015variational}
presents VAEs for anomaly detection, but does so on MNIST and security datasets.

\cite{higgins2016beta} presents $\beta$-VAEs that weights VAE loss terms, but
applies it to datasets much smaller than here. \cite{davies2013searching}
introduce the idea of performing anomaly detection on lunar surface data, only
considering human, manual review, however. \cite{moseley2020unsupervised} use a
VAE for understanding thermal measurements and the thermophysical dynamics of
the lunar surface. \cite{https://doi.org/10.48550/arxiv.1812.08681} presents a
report and review on technosignature detection, recommending automated data
processing methods as presented here. \cite{angerhausen2019machine} present a
proposal for technosignature detection using self-supervised learning.
\cite{lesnikowski2020unsupervised} is similar to this current approach of
self-supervised detection on lunar surface imagery, but the amount of data
inferenced and trained on are both about three orders-of-magnitude smaller.
\cite{lesnikowski2022neural} is an abstract for similar methods as here, but is
only a proposal for results. \cite{chickles2021applications} presents a VAE for
anomalies in astrophysical data, but focuses on time-series, rather than image
data. 

Here we develop the first, self-supervised learning approach that avoids the
need for any labelled train data, while promising to find scientifically
interesting and strategically relevant samples in ultra-large datasets. We
validate our approach by providing metrics on known anomalous samples and a
qualitative review of top returned samples.

In brief we present the following list of contributions:
\begin{itemize}
    \item The first demonstration of the effectiveness of a self-supervised
    approach using a deep generative model towards automated, agnostic retrieval
    of scientifically interesting samples from an ultra-large planetary remote
    sensing dataset.
    \item A codebase that may be extended to work on other large scientific and
    remote sensing datasets.
    \item A new lunar surface imagery dataset with pixel-accurate labeling of
    known human landing sites and a number of geologic features.  
\end{itemize}

\begin{figure}[t]
    \centering
    \includegraphics[width=0.48\textwidth]{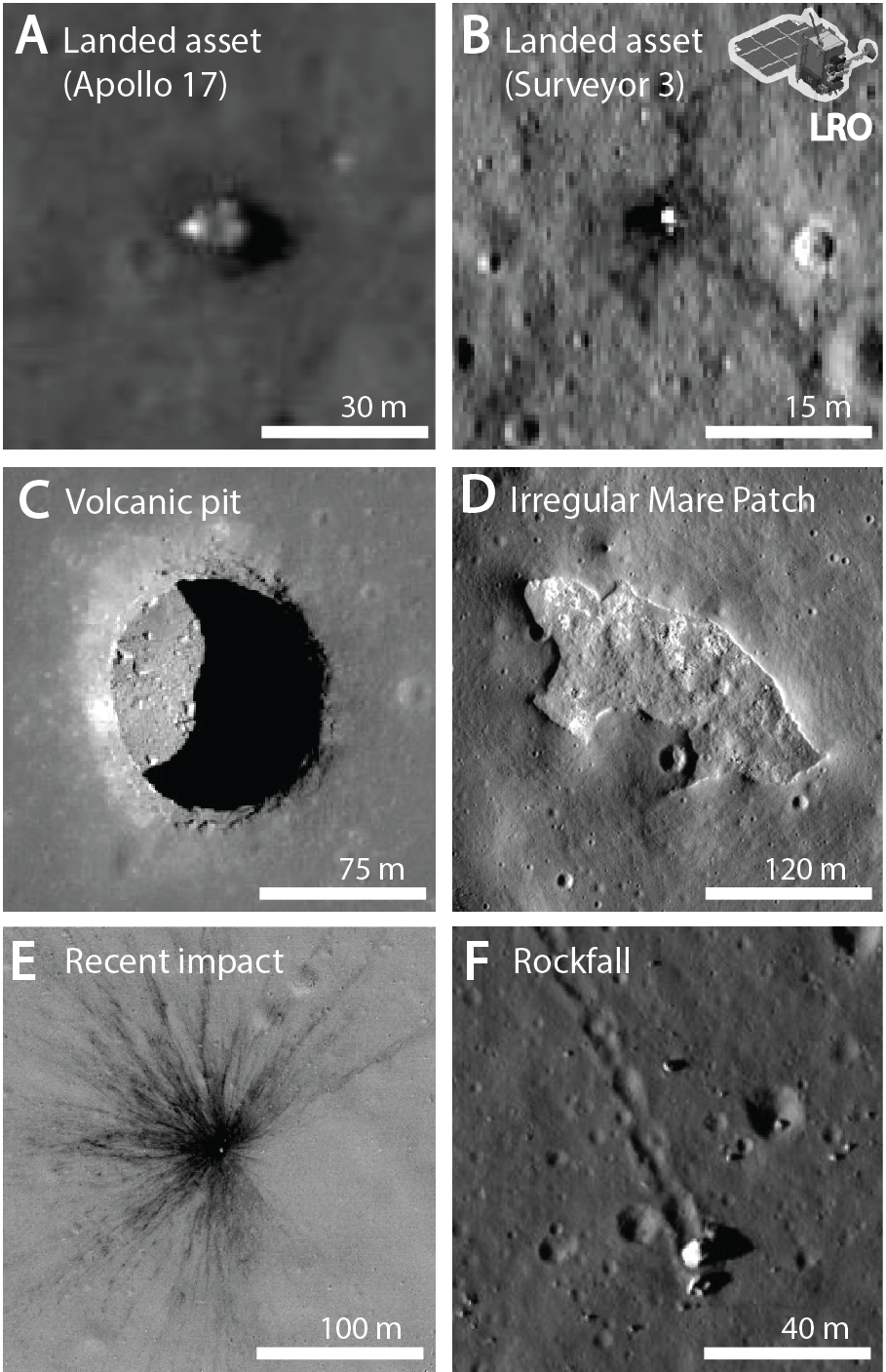}
    \caption{Notable human-made and geologic features of high scientific and/or
    strategic value considered in this work, as imaged by LRO NAC (LRO shown as
    inset): A \& B) landed assets, C) volcanic pits (skylights), D) irregular
    mare patches (IMPs), E) fresh impact craters, F) rockfalls (displaced
    boulder with associated track). Raw image credits to NASA/LROC/GSFC/ASU.
  }
    \label{fig:fig0}
\end{figure} 

\section{Material and Methods}
\subsection{Experimental Setup}
For the verification of our method we produce both quantitative metrics on a
test set of labelled known scientifically interesting examples, and a
qualitative review that our method preferentially retrieves scientifically
interesting examples. For known scientifically interesting examples we take
images that contain the Apollo 12, 15, 16 descent stage Lunar Modules and the
smaller Surveyor 3 probe as shown in Figure \ref{fig:fig0}. In addition, we
include relevant geologic examples, specifically irregular mare patches (IMPs),
fresh impact craters, rockfalls (displaced boulders with an associated track),
and volcanic pits (skylights). Here, a `fresh' impact crater is defined as a
crater that is geologically young, with pronounced ejecta rays and a rocky
interior (see Figure \ref{fig:fig0}). For computing quantitative metrics we
generate a dataset of lunar surface imagery with pixel-accurate labeling of
these sites. We optimize our method by querying its performance on a validation
set. Once optimization is done, we freeze the data, algorithm, and model, and
report the performance on our test sets, to minimize any test set information
leakage issues. To provide quantitative metrics for our method, we frame our
problem as a two-class detection problem. The metrics that we use are the area
under the precision-recall curve and the relative improvement of this area over
a random baseline. We use one cloud NVIDIA A100 GPU for training and one local
2070 RTX Super GPU for validation and test set evaluation.

\subsection{Data}
We use data provided by the official NAC data site at \url{lroc.sese.asu.edu}.
We download PTIFF (pyramid Tagged Image File Format) data which are
full-resolution, calibrated 8 bit versions of the original 12 bit science
images. A PTIFF image contains a number of child TIFF images of varying
resolutions, of which we use the finest resolution available. A generic
finest-resolution TIFF image is 0.5 to 2 meters per pixel resolution, 52 K x 5 K
pixels, and 40 MB in size. We partition each of these raw TIFF images into sets
of 64 x 64 pixel processed images. For our train set we collect a globally
random collection of raw images, which generates a total of 52 million train
patches. For our validation and test sets, we collect all available raw images
of the Apollo 12, 15, 16 and Surveyor 3 landing sites, as well as a number of
raw images containing known geologically interesting features, such as irregular
mare patches (IMPs), volcanic pits, and fresh impact craters. We filter out
low-quality images such as high solar incidence (extensive shadows and poor
signal-to-noise ratios) or emission angle (oblique and distorted geometry), low
resolution, or mission error flags, based on values present in the TIFF file
metadata. Our validation set then consists of the Apollo 15 mission images,
while the test sets consist of the remaining Apollo and Surveyor missions, in
addition to the geological sites. We manually label the patches in our image
that contain the known site of interest as our positive sample, and treat all
other patches as negative samples, generating a total of 11.32 M validation and
test samples. The dataset characteristics that we use for our validation and
test sets are summarized by feature in Table \ref{tab:dataset_characteristics}.
The lunar surface imagery dataset that we contribute consists of human-generated
and verified pixel location labels of all positive samples for our validation
and test sets, together with the parent LROC image id these positive samples are
located in, all grouped by feature type, together with code to dynamically
generate a training set. We make our codebase and this dataset available at
\url{github.com/lesnikow/jstars-automated-discovery}. 

\begin{table}[ht]
\centering
\caption{Number of positives, negatives, and dataset splits for each feature.}
\begin{tabular}{lrrr}
\toprule
\textbf{Feature} & \textbf{Positives} & \textbf{Negatives} & \multicolumn{1}{c}{\textbf{Split}} \\
\midrule
\textbf{Craters}        & 12    & 520,756   & Test \\
\textbf{IMP}            & 4     & 257,852   & Test \\
\textbf{Pits}           & 7     & 419,641   & Test \\
\textbf{Rockfalls}      & 31    & 1,366,353 & Test \\
\textbf{Apollo 12}      & 40    & 2,548,185 & Test \\
\textbf{Apollo 15}      & 40    & 2,498,072 & Val \\
\textbf{Apollo 16}      & 18    & 1,224,798 & Test \\
\textbf{Surveyor 3}     & 39    & 2,483,760 & Test \\
\bottomrule
\textbf{Totals}         & 191	& 11,319,417& --- \\
\end{tabular}
\vspace{0.25cm}
\label{tab:dataset_characteristics}
\end{table}

The Apollo 12 and Surveyor 3 sites are covered by the same NAC parent images,
and so have almost the same number of positive and negative patches. The difference by
one in positive patches between these two sites is due to the fact that the
Surveyor 3 site has one positive patch filtered out due to being too close to
the edge of the NAC image, which contain a black border (masked pixels). 

\subsection{Algorithm}
We train a convolutional variational autoencoder on the train set described
above. At test-time we compare the output of this autoencoder with its input
image, together with how it was generated, to produce an anomaly score for our
test sample. This anomaly score $a(x)$ is computed as a weighted sum of the
squared $L^2$ norm between our input image $x$ and its reconstruction $\hat{x}$,
together with the squared $L^2$ norm of the $\mu$ latent code of the image.
Specifically, our anomaly score $a(x)$ for input image $x$ is
\begin{equation}
    a(x) = \lVert x - \hat{x} \rVert_{2}^{2} + \lambda \cdot \lVert \mu \rVert_{2}^{2},
\end{equation} 
where $\lambda$ is a hyperparameter that was chosen by tuning on our validation
set. The motivation of this second component is that during training, the VAE
has, as part of its loss, a distribution matching term between mapped codes and
a prior Gaussian distribution. Under the assumption that codes far from the
distribution mean are anomalous in pixel space as well, we include this
distribution loss in the anomaly score as a component weighed against the
reconstruction loss. We obtain test-time anomaly scores by a single inference
pass of a trained model on test samples.

We choose no single anomaly score that separates an in-distribution from an
out-of-distribution or anomalous sample. An optimal cut-off score might vary
from application to application, e.g., whether one is interested in
technological anomalies (i.e., landers) or natural anomalies (i.e., pits,
craters, etc.). This choice should also be influenced by the relative costs
incurred by false positives versus false negatives. Hence we stipulate no single
cut-off, but instead show the trade-off in model performance as we sweep through
all possible decision thresholds in the precision-recall curves of
Figure~\ref{fig:pr_curves}.

We use regularization techniques to combat overfitting to the validation set,
and a diverse test set to measure model generalization. For regularization
methods, we have batch-normalization layers in our encoders and decoders,
adaptive learning rates and momentum terms in our Adam network optimizer,
stochastic mini-batch selection in our SGD-based optimizer, and a large, diverse
training set. These regularization methods have been well-tested at combating,
among other issues, validation set overfitting. For a test set, we use a
collection of diverse sites that were never trained or validated on. The metrics
on these test sets were computed just once at the end of our experiments, after
training and tuning were finished. These test sets provide unbiased estimates of
our method’s generalization to new, unseen data.

Our motivation in using a VAE rather than e.g. a generative adversarial network
(GAN) is that VAEs have a natural anomaly score to use, namely some variant of
their reconstruction error between input and output images. On the other hand,
GANs do not, since they typically generate images from noise vectors, and hence
they require more work, such as additional density estimation methods, to use
for anomaly detection. VAEs are also typically easier to train to convergence
than GANs.

\subsection{Model}
Our VAE has an encoder with four convolutional layers and one fully connected
layer. In its middle bottleneck layers, this model has its encoder output fully
connected to one set of $d = 256$ nodes, which is fully connected to $\mu$ mean
and $\sigma$ deviation nodes, each set of size $d$, which are in turn fully
connected to two fully connected layers, again each of size $d$. This bottleneck
output is fed into a decoder, symmetric to our encoder, see Figure
\ref{fig:network_diagram} for a diagram. There are batch-norm and ReLU layers in
between each convolutional layer of our encoder and decoder, batch-norm layers
after the fully connected layers in our bottleneck, and a VAE reparameterization
layer after our $\mu$ and $\sigma$ nodes. All together our model has thirty-two
layers. Our model is trained with a standard VAE loss, but with the
reconstruction loss as the $L^1$ distance between input and reconstruction, and
with the distribution loss weighted, as in a $\beta$-VAE, as $\beta$ times the
reconstruction loss, for $\beta = \frac{1}{4}$. These two noted choices were
found through hyperparameter optimization. We train with a batch size of 8192
with the Adam optimizer at a learning rate of 1e-3 for a total of five training
epochs, exhausting our train budget.

\section{Results}
\begin{figure*}[h!]
    \centering
    \includegraphics[width=2\columnwidth]{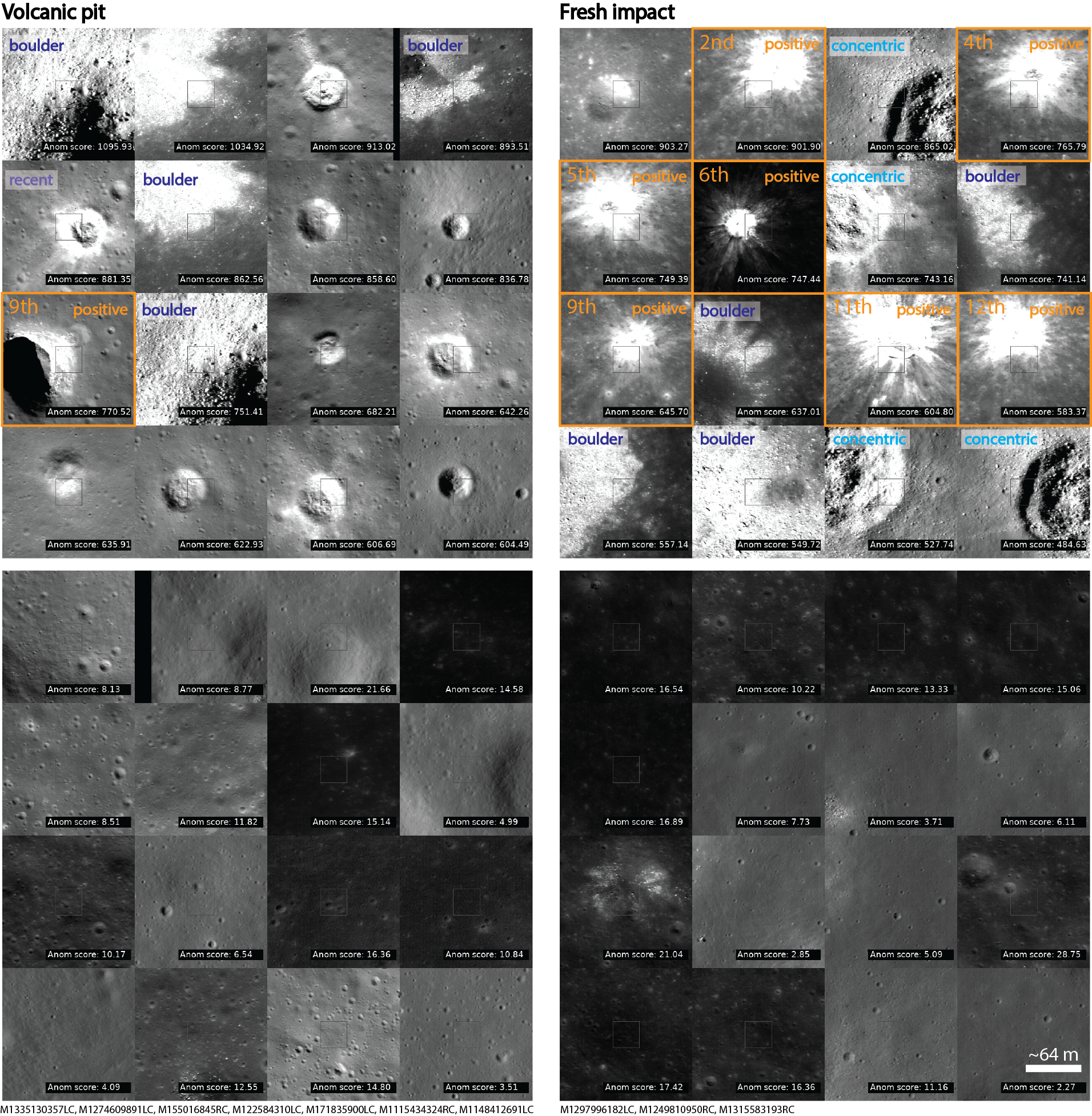}
    \caption{Top: most anomalous 16 patches for pit and crater images, Bottom:
    random 16 patches for pit and crater images. Positive features indicated in
    orange (positive); recent craters (not labelled as positives) indicated in
    violet (\textit{recent}); concentric craters indicated in blue
    (\textit{concentric}); boulder fields and rocky craters indicated in dark
    blue (\textit{boulder}). The model's 64 x 64 pixel input window is drawn as
    the smaller interior rectangle in each patch, while surrounding pixels
    outside of this input window are added in this and other figures for
    interpretable context. Note that some features are detected multiple times,
    i.e., by different windows. Raw image credits to NASA/LROC/GSFC/ASU. }
    \label{fig:most_anomalous_vs_random_pit+crater}
\end{figure*}

We first provide a comparison of the top-scoring patches versus randomly
selected patches, to provide a qualitative comparison of our method to random
manual review. The top sixteen positive patches for our volcanic pit and fresh
crater test sets, along with their anomaly scores, are compared to an equal
number of random patches and scores from the pit and crater test set in Figure
\ref{fig:most_anomalous_vs_random_pit+crater}. The most anomalous patches show
fresh, bright, boulder-rich parts of the lunar surface, with one positive pit
sample (among 419,648 searched patches) and seven positive crater samples (among
520,768 searched patches) appearing among the top sixteen candidates of searched
patches. All randomly chosen patches show dark, smooth, and feature-less parts
of the lunar surface, being representative of the Moon's overall appearance. We
note that processing of the entire test set with more than 11 million patches
took only about about one-half hour in a single consumer-level GPU, and is
highly parallelizable across multiple GPUs. Figure
\ref{fig:all_positives_crater+pit} shows all positive pit and fresh crater
features with their respective anomaly scores.

\begin{figure}[t]
    \centering
    \subfigure[]{\includegraphics[width=0.48\columnwidth]{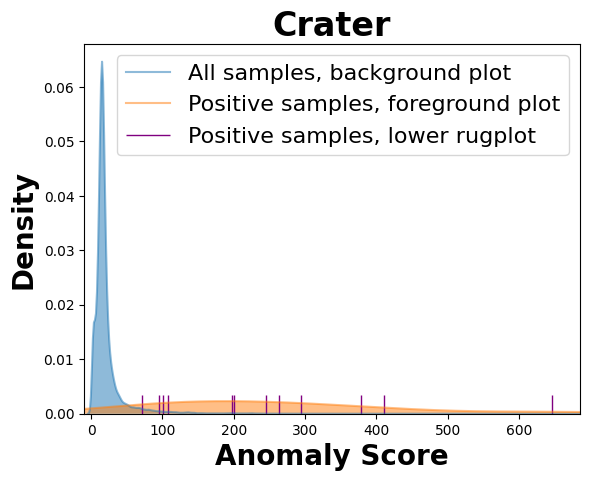}}
    \subfigure[]{\includegraphics[width=0.48\columnwidth]{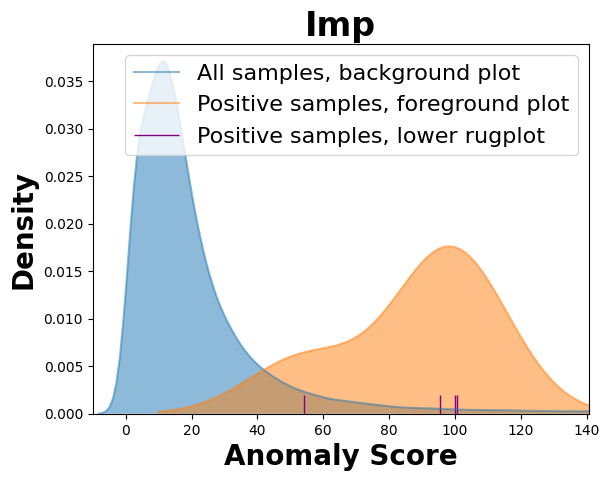}}
    \subfigure[]{\includegraphics[width=0.48\columnwidth]{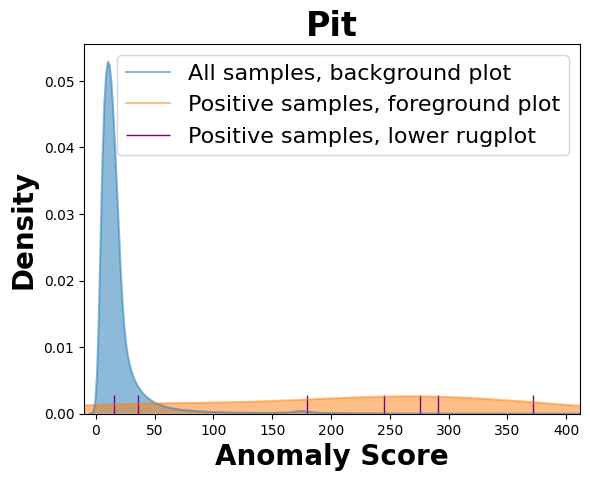}}
    \subfigure[]{\includegraphics[width=0.48\columnwidth]{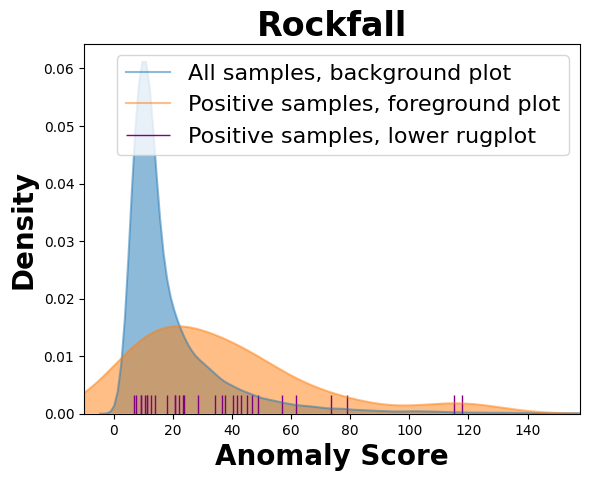}}
    \subfigure[]{\includegraphics[width=0.48\columnwidth]{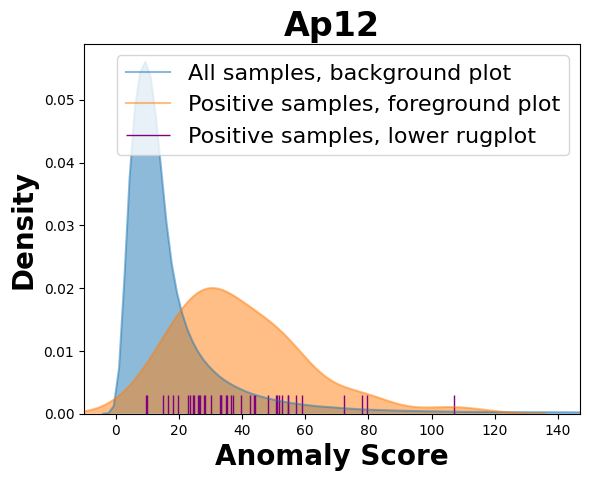}}
    \subfigure[]{\includegraphics[width=0.48\columnwidth]{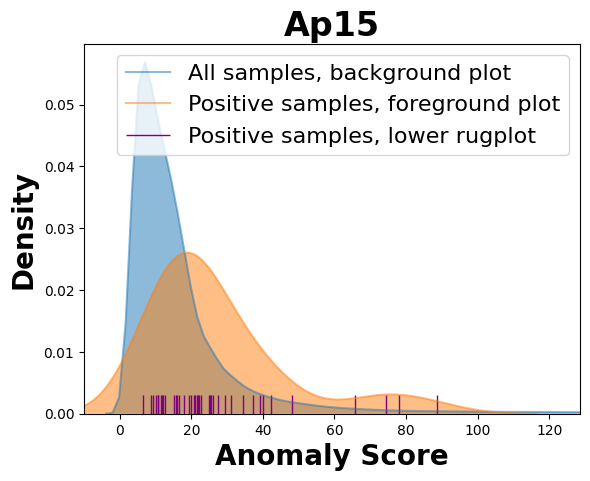}}
    \subfigure[]{\includegraphics[width=0.48\columnwidth]{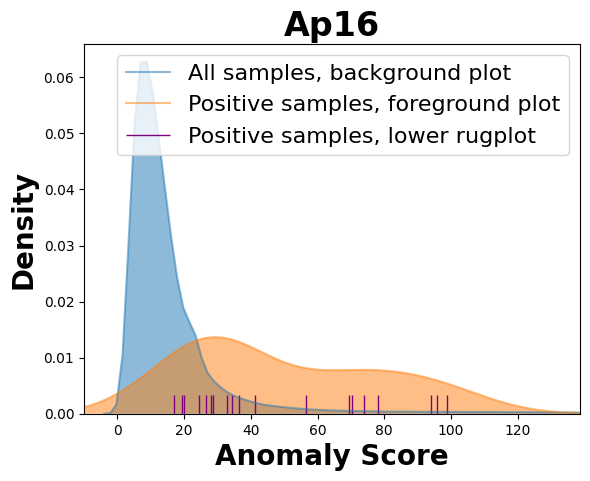}}
    \subfigure[]{\includegraphics[width=0.48\columnwidth]{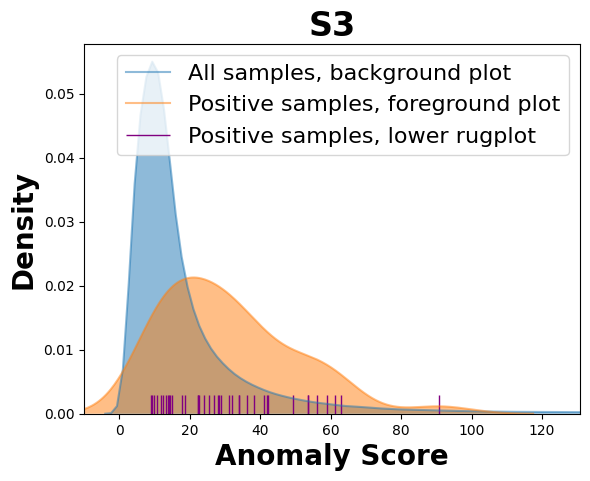}}
    \caption{Bottom rugplots show scores for positive images, while the
    background KDE plots show all image scores.
    (a) craters, (b) irregular mare patches, (c) volcanic pits, (d) rockfalls,
    (e) Apollo 12 (f) Apollo 15 (g) Apollo 16 (h) Surveyor 3.  }
    \label{fig:kde_plots}
\end{figure}

We next investigate the anomaly scores of positive patches versus all patches.
Figure \ref{fig:kde_plots} shows the distribution of anomaly scores for all
positive versus all patches for our two test Apollo landing sites. Table
\ref{tab:ks_statistics} provides a table of two-sample Kolmogorov-Smirnov test
statistics to test whether the empirical distributions of model anomaly scores
between positive and all patches is different in a statistically significant
manner at an $\alpha = 0.05$ significance level. The model's anomaly score
distributions for positive examples are statistically significantly different
from all samples for all test classes. 

\begin{figure}[h!t]
    \centering
    \subfigure[]{\includegraphics[width=0.48\columnwidth]{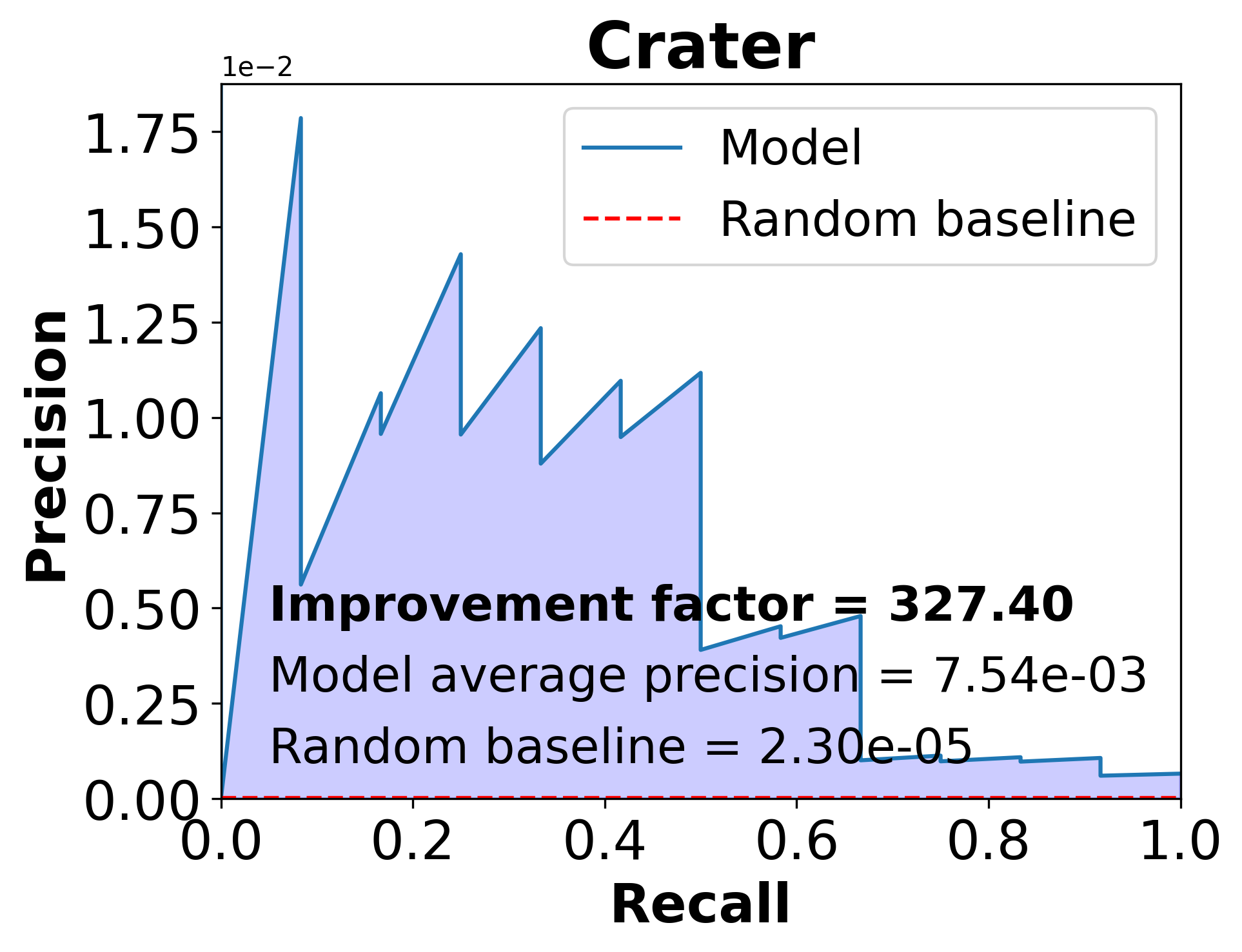}} 
    \subfigure[]{\includegraphics[width=0.48\columnwidth]{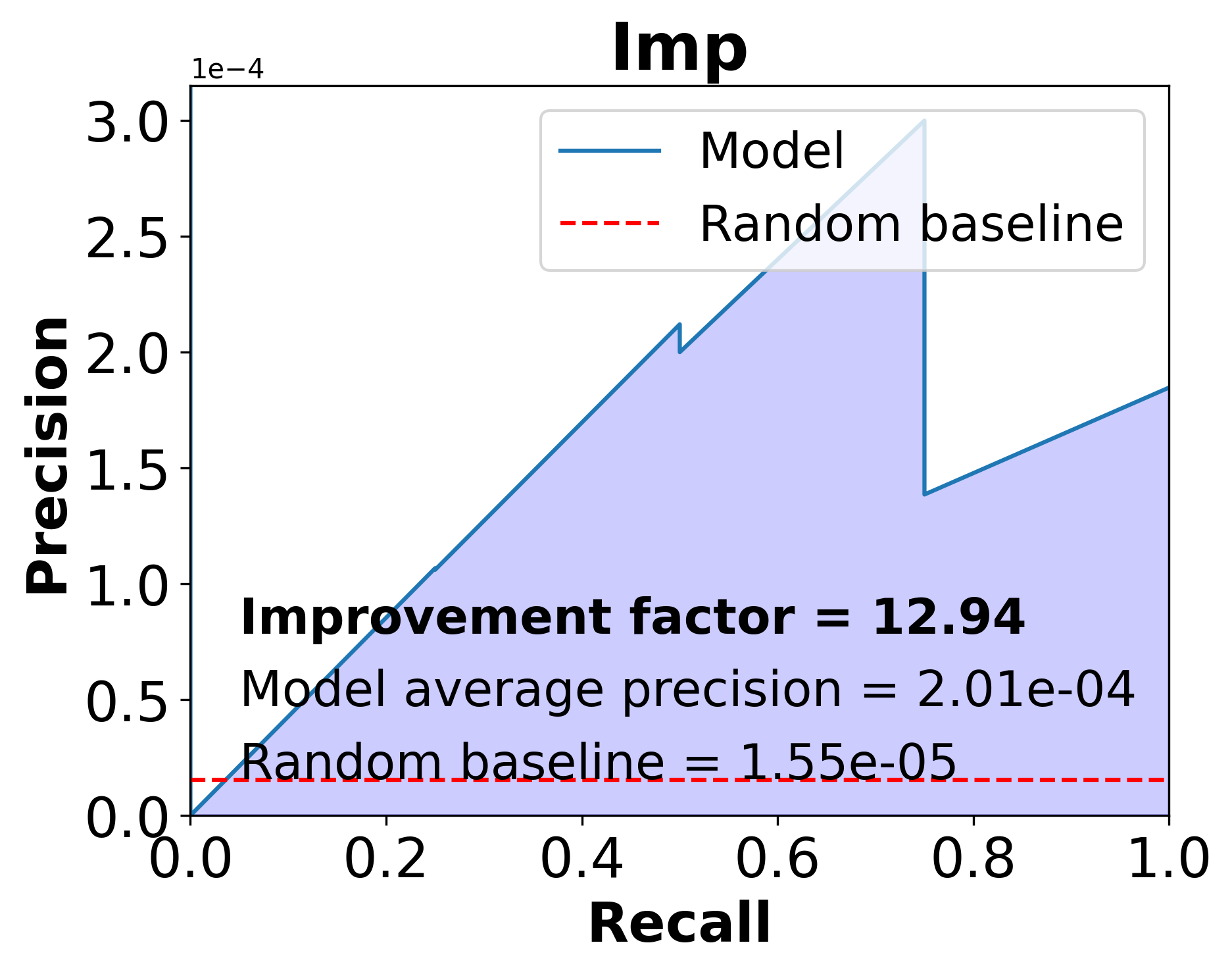}} 
    \subfigure[]{\includegraphics[width=0.48\columnwidth]{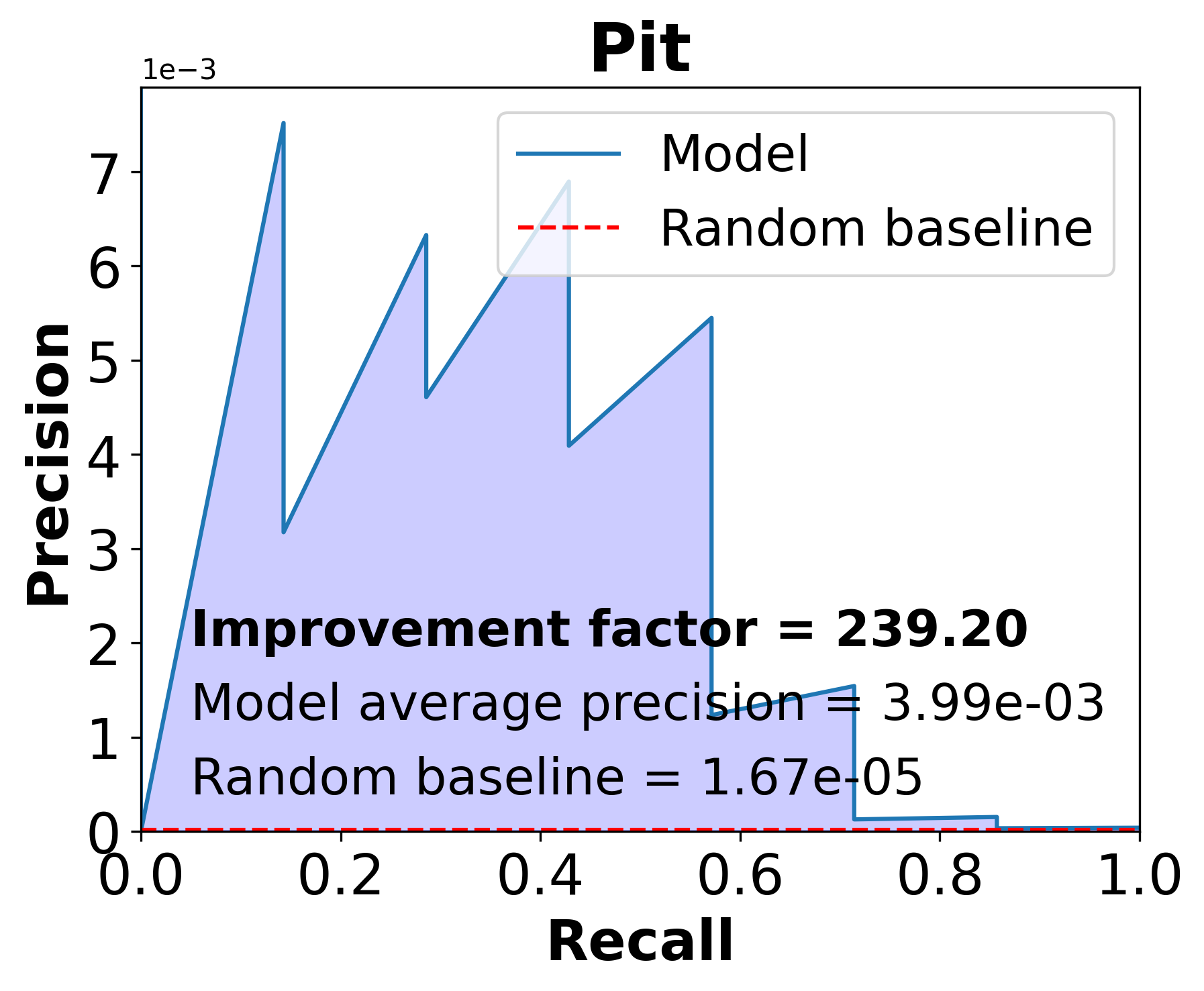}} 
    \subfigure[]{\includegraphics[width=0.48\columnwidth]{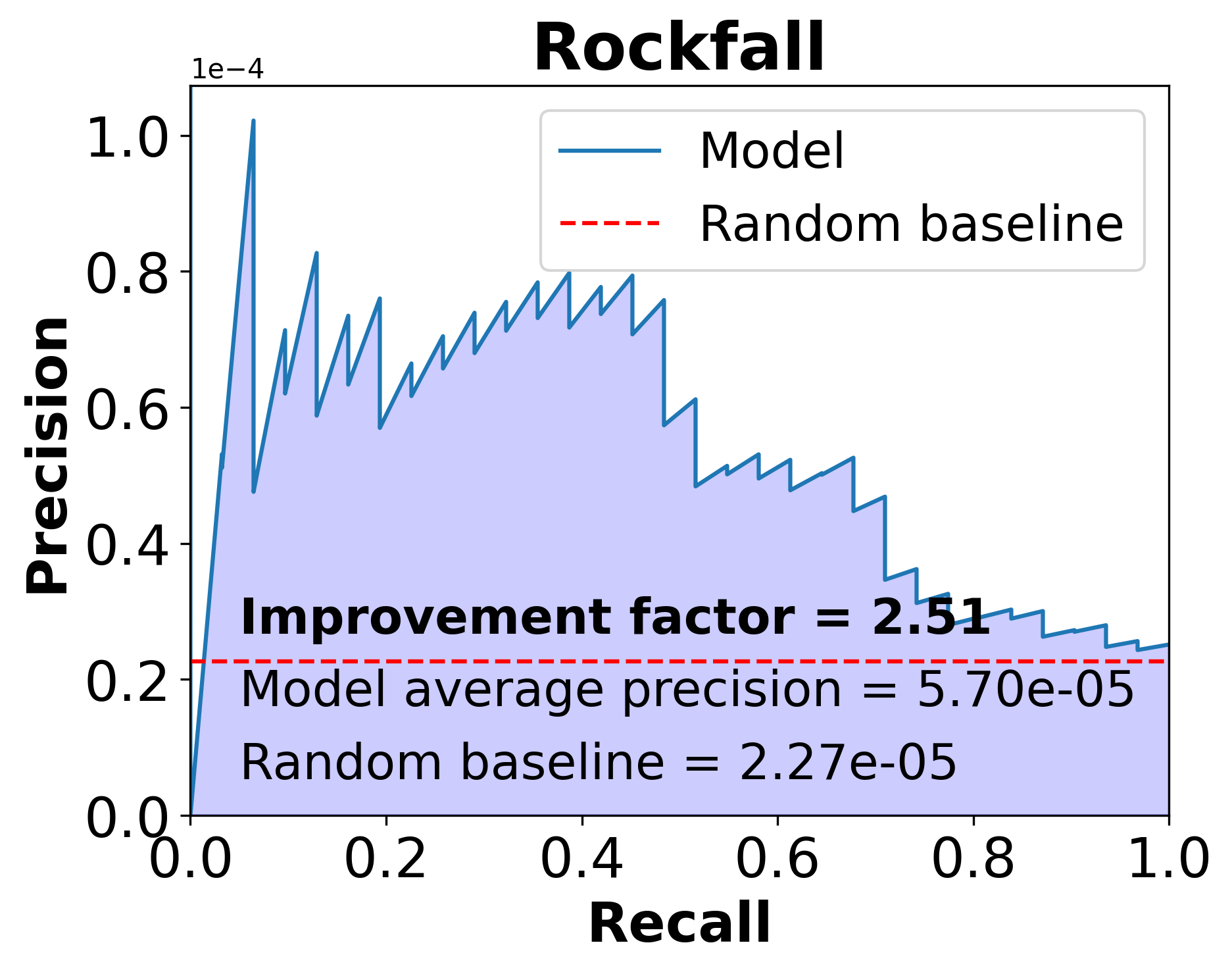}} 
    \subfigure[]{\includegraphics[width=0.48\columnwidth]{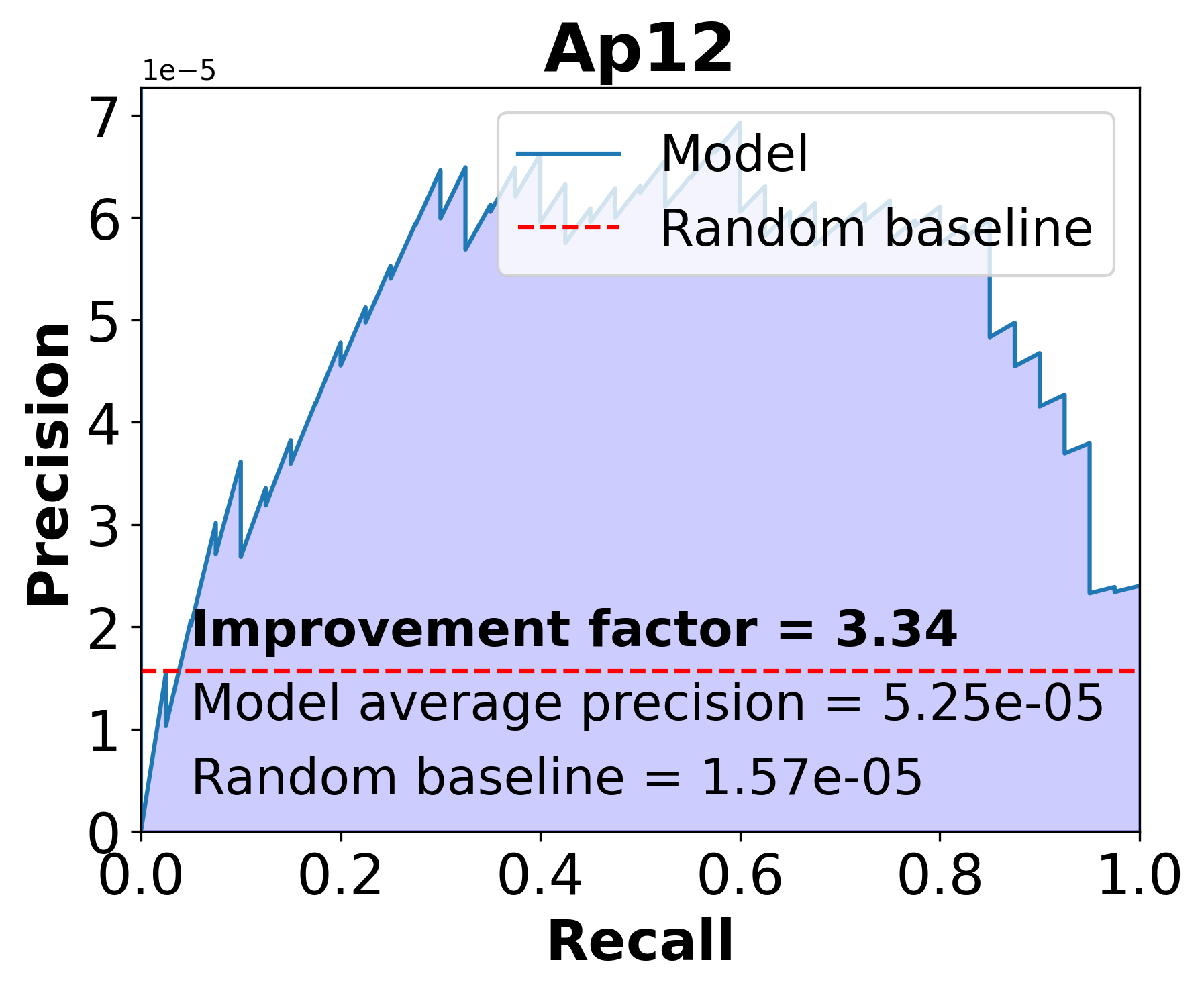}} 
    \subfigure[]{\includegraphics[width=0.48\columnwidth]{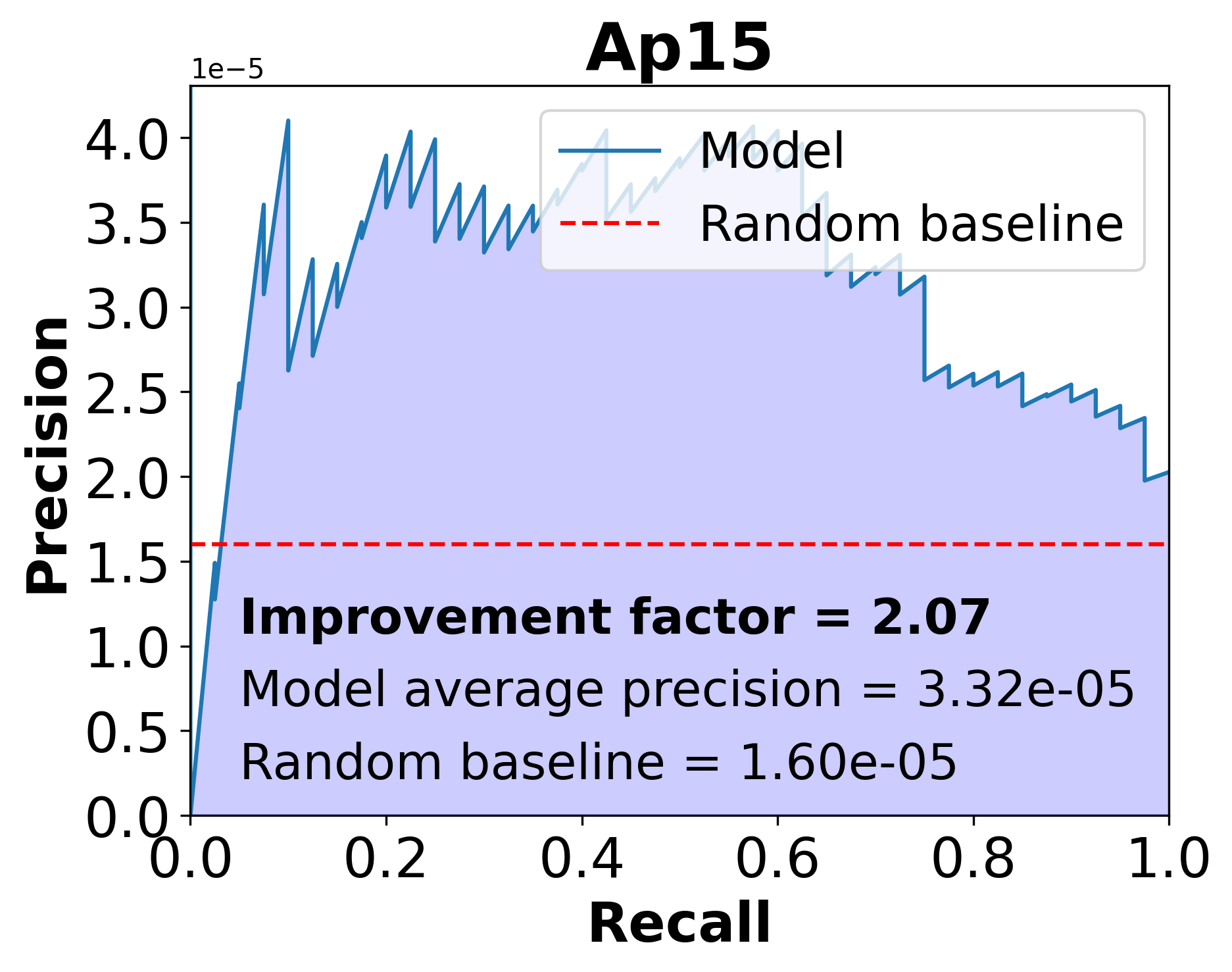}} 
    \subfigure[]{\includegraphics[width=0.48\columnwidth]{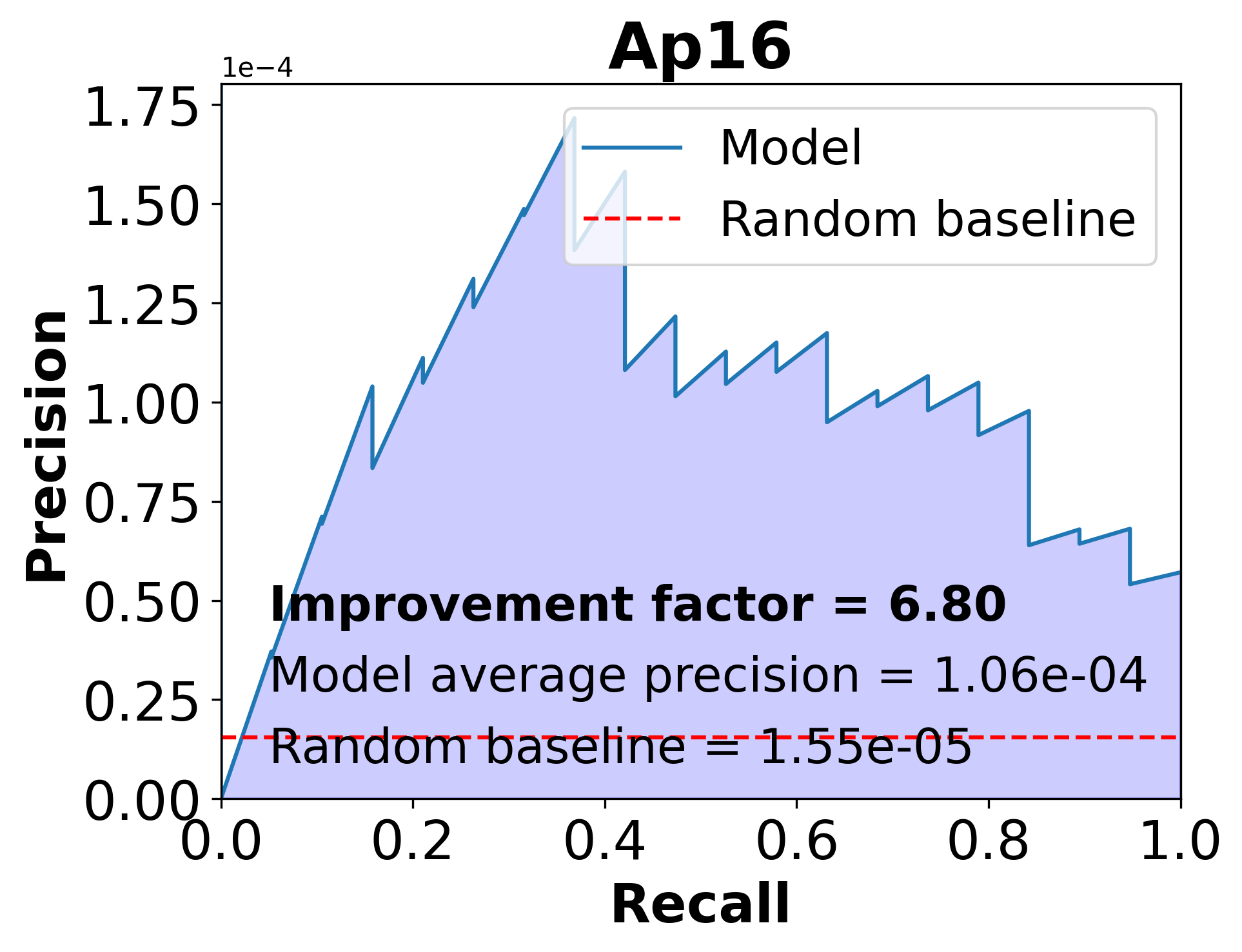}} 
    \subfigure[]{\includegraphics[width=0.48\columnwidth]{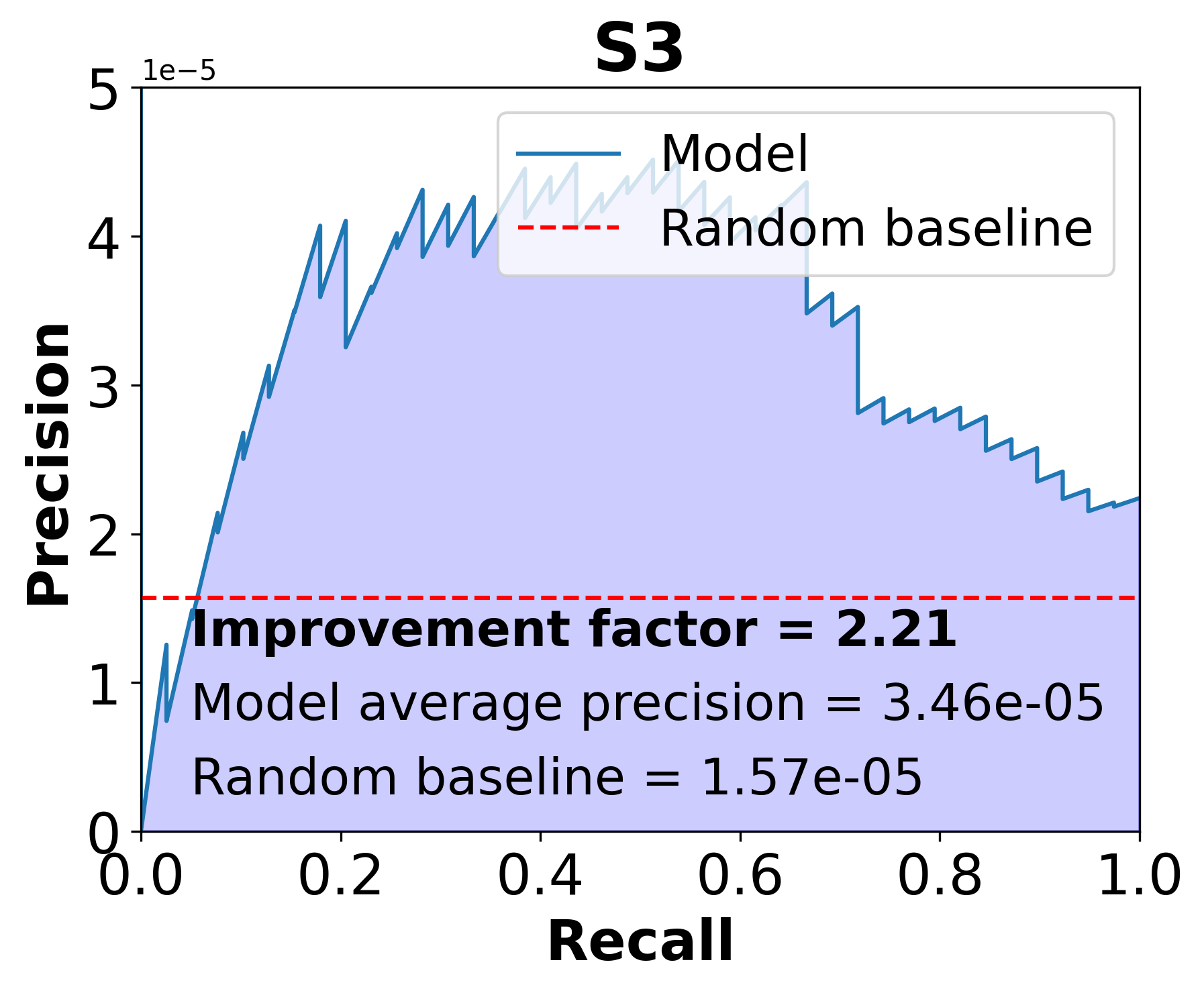}} 
    \caption{Precision-recall curves for validation and test sets.
    Y-axes are scaled separately for each plot for improved readability.
    (a) craters, (b) irregular mare patches, (c) volcanic pits, (d) rockfalls,
    (e) Apollo 12 (f) Apollo 15 (g) Apollo 16 (h) Surveyor 3.  }
    \label{fig:pr_curves}
\end{figure}

We plot precision-recall curves and calculate average precisions to provide
quantitative performance metrics across the range of the model's predicted
samples. Figure \ref{fig:pr_curves} shows these curves, together with the random
prediction baseline curves for these missions and features. As there are
significantly more negative samples than positives, we report the relative
performance gain of our method versus a random baseline. These relative
performance improvement factors, as well as a tabular summary of single-number
average precision metrics, are provided in Table \ref{tab:ap_model_vs_baseline}.
The precision-recall curves underline how the model significantly outperforms a
random baseline, while performing particularly well for the fresh crater, IMP,
and volcanic pit classes. Figure \ref{fig:pr_curves} further indicates that some
of the highest scoring anomalies are (apparent) negative patches, an observation
that is further discussed below.

\begin{figure*}[h!t]
\centering
\includegraphics[width=0.8\textwidth]{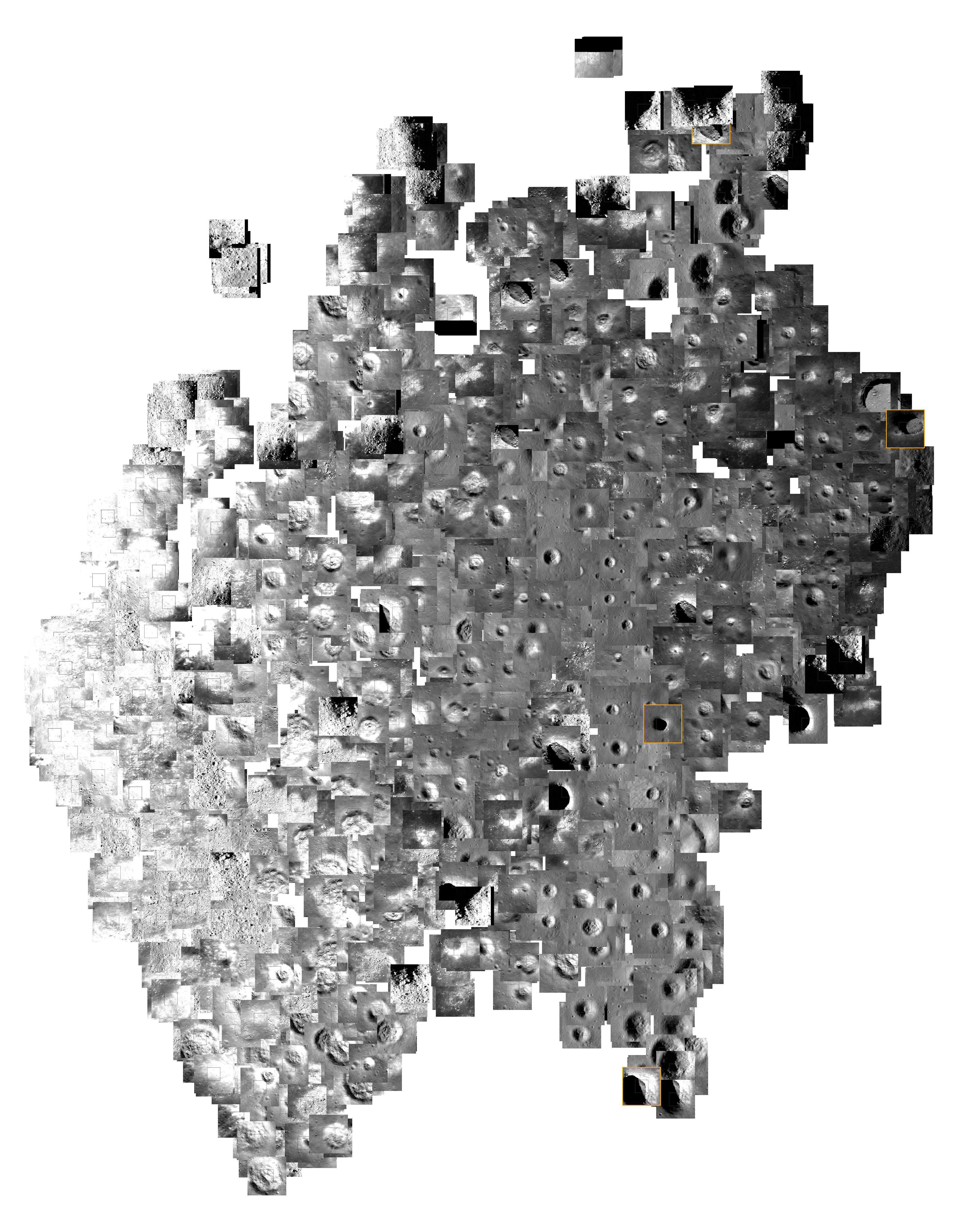}
\caption{t-SNE plot of top 2048 anomalous samples from the pit test set. See
Figure \ref{fig:fig0} for a volcanic pit example. Samples labelled as true
positives are colored here with an orange border. Note that pits are often
larger than the model's 64 x 64 pixel model window size, so that while each pit
only has one labelled positive sample which is colored here, the same pit may
appear in multiple plotted samples, due to multiple, different model input
windows covering the same pit. Raw image credits to NASA/LROC/GSFC/ASU.
}
\label{fig:tsne_pit}
\end{figure*}

We provide a t-SNE plot of the top anomalous images of our crater and pit test
sets in Figures \ref{fig:tsne_pit}, \ref{fig:tsne_crater} respectively. These
plots are used to provide a low-dimensional visualization of a
high-dimensional dataset. A t-SNE plot is optimized to plot data samples that
are similar in their native high-dimensional space to be close in two dimensions
, while those that are dissimilar in high-dimensional space are free to to
be plotted further away from each other. In practice, t-SNE visualizations for
images tend to cluster similar-looking images into discernible clusters in an
automated way. Some of these clusters may be human-interpretable and
scientifically interesting. The volcanic pit t-SNE plot for the 2048 most
anomalous patches (Figure \ref{fig:tsne_pit}) shows gradients in reflectance
from left (bright) to right (dark) and feature size from top left (larger) to
bottom right (smaller). The left hemisphere of the plot is occupied by
boulder-rich crater ejecta blankets without distinct shadows, whereas the right
hemisphere mostly consists of small, intermediately old impact craters with
partially shadowed slopes. We note that most of the volcanic pit patches are
located in the center and right hemisphere of the t-SNE plot, as pits tend to be
circular depressions with shadowed floors. In the t-SNE plots here, images with
higher anomaly scores are foreground to images with lower anomaly scores in
order to give a visual sense of the relative ordering of images by anomaly
score.

\begin{table}[ht]
    \centering
    \caption{Two sample Kolmogorov–Smirnov test results by feature type.}
    \begin{tabular}{lrrl}
        \toprule
        Feature &  K-S test statistic & p-value & Significant? \\
        \midrule
        Crater &  0.964967 &  3.941793e-10 & True \\
           IMP &  0.915987 &  2.432050e-02 & True \\
           Pit &  0.763843 &  5.670249e-04 & True \\
      Rockfall &  0.385228 &  2.019546e-04 & True \\
     Apollo 12 &  0.626816 &  4.472798e-14 & True \\
     Apollo 15 &  0.374041 &  2.755836e-05 & True \\
     Apollo 16 &  0.731476 &  2.958198e-09 & True \\
    Surveyor 3 &  0.426762 &  1.353984e-06 & True \\
    \bottomrule
\end{tabular}
    \vspace{0.25cm}
    \label{tab:ks_statistics}
\end{table}

\section{Discussion}  

In qualitatively comparing top scoring to random images in Figure
\ref{fig:most_anomalous_vs_random_pit+crater}, we observe significantly more
train-time unknown but notable geological features such as fresh craters,
boulder fields, volcanic pits, and other features. This is done by our automated
method at a speed of ten seconds for a full 52 K x 5 K pixel image. Assuming a
human review rate of one 64 x 64 pixel patch per second for a review of
fine-scale features, a full image would take 1 second / patch x 64 K patches /
image = 64 K seconds / image ~= 17.8 hours / image. This translates into a
time-efficiency gain of 17.8 hours / 10 seconds ${\sim} 6.4 \times 10^3$. This
time-efficiency gain would be in addition to the human labor and opportunity
costs recovered from automation, along with potentially decreasing human
performance due to fatigue that automated methods do not suffer from. In Figure
\ref{fig:kde_plots} we note a (statistically significant) rightward shift of the
positive sample score distribution relative to the negative sample score
distribution. We interpret this to mean that our method preferentially gives
higher anomaly scores to positive samples than to negative background samples.
Similarly, in the precision-recall curves of Figure \ref{fig:pr_curves} we
observe improved performance over a range of thresholds for our method compared
to the random baselines. The average precision improvement factors for our
method on our test set range from 2.20 for the smaller Surveyor 3, 6.84 for the
larger Apollo 16 site, 239.20 for pits, and 327.40 for craters, as shown in
Figure \ref{fig:kde_plots}. Our model performs best in recognizing the fresh
crater, IMP, and volcanic pit classes, but not as well in recognizing the other
classes. We interpret this to be representative of the visual, geomorphic
uniqueness of the features themselves, where fresh craters, IMPs, and volcanic
pits are highly unique in terms of shape and texture (see Figure
\ref{fig:fig0}). For example, IMPs and volcanic pits feature highly distinct
edges (i.e., abrupt bright to dark contrasts), while fresh craters feature
rocky, boulder-rich (bright) interiors and pronounced ejecta rays. In contrast,
rockfalls and landed hardware tend to resemble regular boulders - which are
relatively rare on the lunar surface, but not as rare as rockfalls and landed
assets, for example. 

As indicated by Figure \ref{fig:pr_curves}, many top-scoring patches are
(apparent) negatives. While the respective patches have been officially labelled
as `negatives', e.g. as `not a volcanic pit' in the volcanic pit test set, they
might not be non-anomalous in general. Negative patches can still contain highly
anomalous features that are not captured by the respective test class and skew
the test results, such as boulder fields, recent (vs fresh) craters, and
concentric craters, as shown in Figure
\ref{fig:most_anomalous_vs_random_pit+crater}. In addition, we note that some
test classes suffer from a human labelling bias, such as the fresh crater class:
impact craters experience steady degradation after their formation, meaning it
is impossible to fully separate `fresh' vs `old' craters in a labelled (binary)
test set. As a result, some of the craters labelled as `not fresh' might achieve
a higher anomaly score than some of the `fresh' craters, skewing the test
results. In other words, all reported testing results underestimate the actual
performance of our model in finding anomalous features on the highly
heterogeneous lunar surface.

We observe that features which are on a similar spatial scale as our model's
window scale (64 x 64 pixels, i.e., between 32 x 32 m and 128 x 128 m) seem to
be retrieved preferentially over features which are either a relatively small
part of our inference window, such as rockfalls and smaller landing hardware
(e.g., Surveyor 3, 3 m across), or features which span across a scale larger
than the inference window, such as very large volcanic pits or impact craters.
We note that targets beyond the current model's spatial sensitivity range could
be recognized with a re-trained model that utilizes smaller or larger windows.
In other words, the anomaly score of any given feature strongly depends on the
relation of its size, the spatial resolution of a given image, and the size of
the model's window. For example, we would expect patches containing relatively
small features like landed hardware and rockfalls to achieve significantly
higher anomaly scores in smaller windows (or better resolved images), as the
ratio between feature-pixels and background-pixels improves, enabling the model
to focus on the subtle differences between the features of interest and the
overall background. Future work will look into the integration of additional
window sizes to cover a wider range of potentially interesting features.

We note that two or more different features only very rarely overlap, given the
relative scarcity of the features considered, their small spatial extent, and
the very large search areas we examine. We also note that our anomaly detector
is feature agnostic\,--\,it merely assigns an anomaly score to a given patch. If
two or more different types of rare features were to be contained in one single
patch we expect the anomaly score would remain high\,--\,or would be higher, given
that the anomaly score is calculated in part as the per-pixel squared
differences between input and output images. The combination of our method
giving consistently higher anomaly scores to known interesting positive sample
together with the vast increase of processing speed that such an automation
gives suggests that such a method would allow researchers to process vast
amounts of data much faster than previously available while finding
needle-in-the-haystack samples (see Figure \ref{fig:anomaly_map}). Such a
capability is particularly useful for global-scale mapping efforts, searching
for known features (such as volcanic pits) as well as searching for completely
unknown features (i.e., science discovery). Based on the currently used hardware
and model, we estimate a full scan of the LRO NAC dataset to take ${\sim}173$
days using a single local 2070 GPU. However this tasks is highly parrallelizable
across multiple GPUs, and we estimate a full scan of the dataset would take
${\sim}21$ days using eight cloud GPUs, or only ${\sim}2$ days using one hundred
cloud GPUs. Besides global-scale mapping and discovery, we note that our
approach is able to accelerate the search for specific features in
time-sensitive scenarios, such as the search for landing/crash sites and
associated debris after the contact to a lander was lost. Such a capability
becomes more and more relevant as the number of (agency and commercial) missions
to the Moon increases.

That our method performs well, on a diversity of test sets, which were not used
at all during training or validation, is strong evidence to us that validation
set overfitting is either minimal or not harmful to the success of our method.
We interpret all of these results together to mean that our proposed methods
can effectively and efficiently retrieve scientifically interesting and
strategically relevant samples, as constructed in our test sets, from
ultra-large planetary science image datasets. This implies to us that this
method can plausibly be extended to produce similar or better results on this
and other large planetary remote sensing datasets. 

\section{Conclusions}
We demonstrate the effectiveness of an automated method for the retrieval of
scientifically interesting and strategically relevant samples from the
ultra-large LRO NAC orbital dataset with more than two million images, for the
first time. Our model is able to systematically and rapidly retrieve rare
features such as volcanic pits, irregular mare patches, fresh craters,
rockfalls, and landed hardware from an overwhelming amount of image patches.
Depending on the feature of interest, our approach provides an average precision
improvement between 2 to 327 times, while being ${>} 10^3$ times faster, than
manual review, making it highly applicable for ultra-large-scale processing. In
addition this automation frees human and scientist reviewers to focus on tasks
requiring relatively more creativity and fluid intelligence, while not being
susceptible to human fatigue from repetitive or tiring tasks. Our approach can
be used to create global-scale maps of anomalous and rare lunar surface
features, search for previously unknown features, and rapidly identify
strategically relevant targets such as spacecraft crash sites in time-sensitive
scenarios. We share the code and machine learning dataset of lunar surface
imagery with labels of known human landing sites and geologic features of
interest that we create to compute quantitative metrics.

\section{Acknowledgements}
This work has been carried out within the framework of the NCCR PlanetS
supported by the Swiss National Science Foundation under grants
PlanetS-TP-2022-SF12, 51NF40\_182901, and 51NF40\_205606. The authors
acknowledge the support of Google Cloud Research Credit Grants 202320389 and 43392715. We would like to thank and acknowledge Divyanshu Singh
Chauhan, Bharathrushab Manthripragada, and the Frontier Development Lab (FDL).

\bibliography{main} 
\bibliographystyle{IEEEtran}

\section{Biography Section}

\begin{IEEEbiography}[{\includegraphics
[width=1in,height=1.25in,clip,
keepaspectratio]{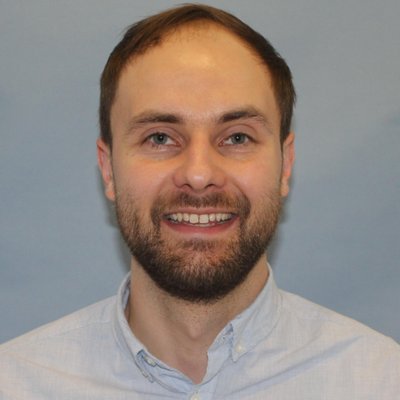}}]
{Adam Lesnikowski} is an independent researcher and consultant based in RI
working on machine learning and artificial intelligence. Previously he worked as
a machine learning scientist and senior software perception engineer at NVIDIA
in Santa Clara, CA, a startup founder, and as a graduate student researcher in
the Ph.D.~program in mathematical logic at U.C. Berkeley in Berkeley, CA. He
also received an A.B.~in mathematics and philosophy from Harvard University in
Cambridge, MA.
\end{IEEEbiography}

\begin{IEEEbiography}[{\includegraphics
[width=1in,height=1.25in,clip,
keepaspectratio]{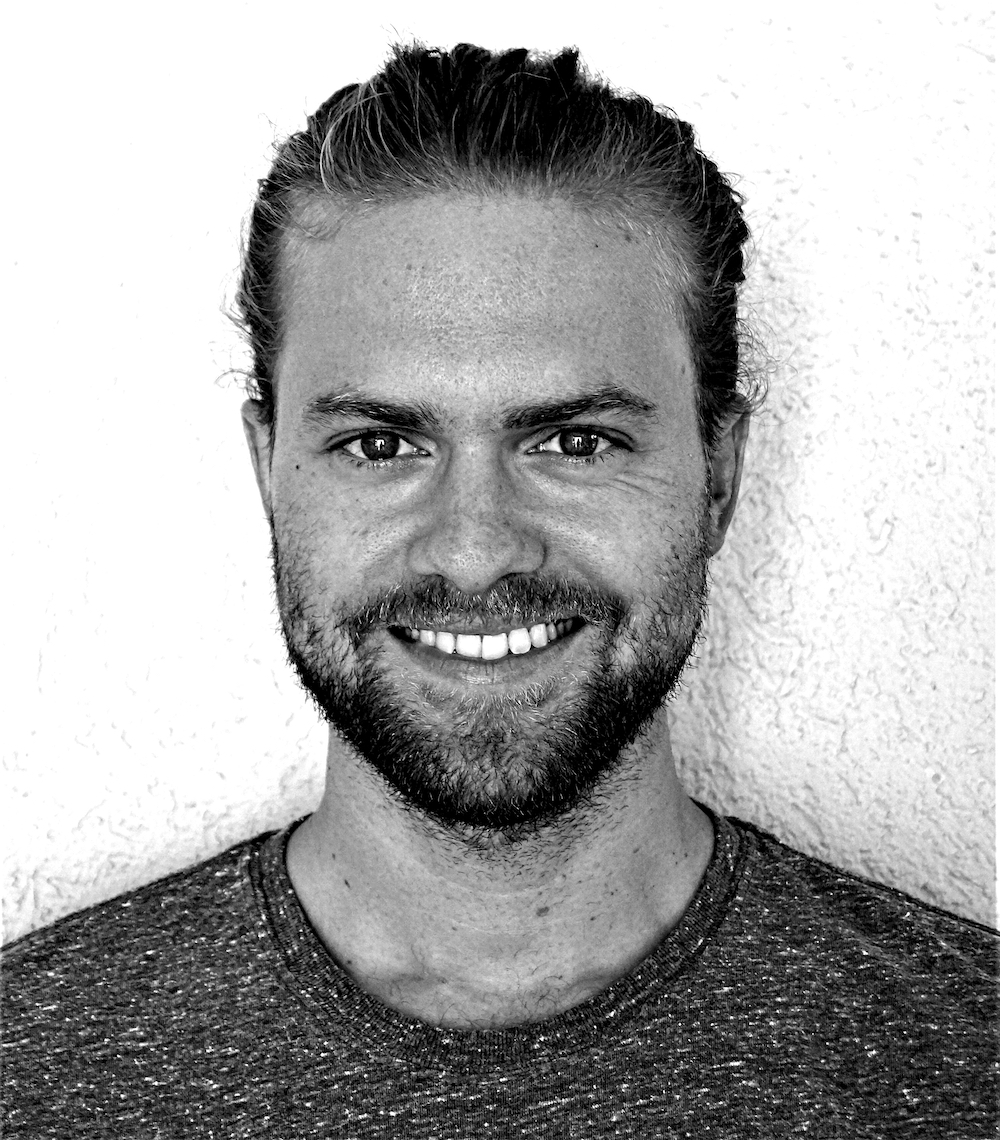}}]{Valentin T. Bickel} received his B.Sc.
in Geosciences from the Technical University of Munich and the Ludwig
Maximilians University Munich, as well as his M.Sc. in Engineering Geology from
ETH Zurich. He received his Dr. sc. ETH degree in Planetary Science from ETH
Zurich and the Max Planck Institute for Solar System Research. He currently
holds a postdoctoral research position at the Center for Space and Habitability,
University of Bern, Switzerland.
\end{IEEEbiography}

\begin{IEEEbiography}[{\includegraphics
[width=1in,height=1.25in,clip, keepaspectratio]{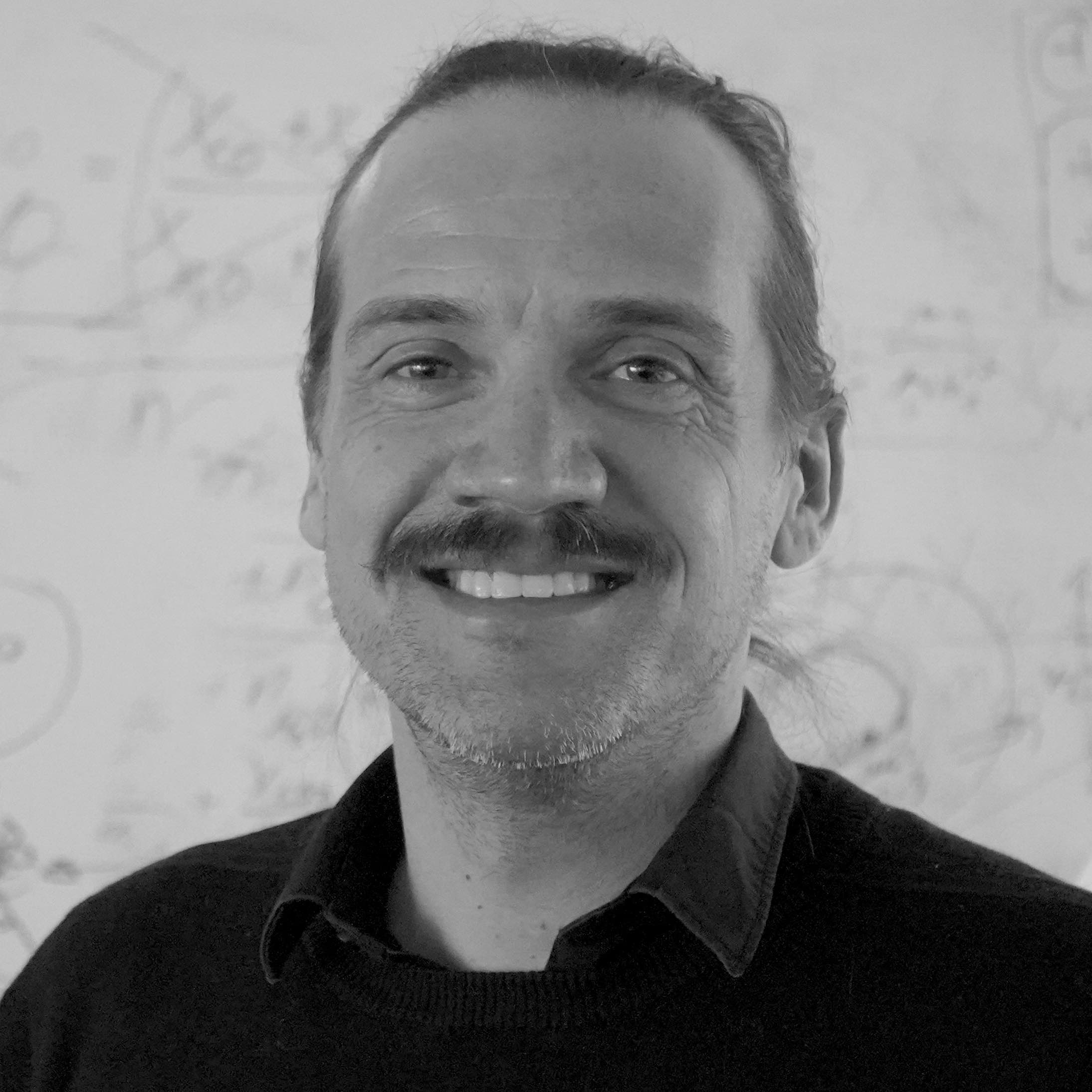}}]{Daniel
Angerhausen} received his Dipl. Phys. from Cologne University and his Dr. rer.
nat. from Stuttgart University. After postdoctoral fellowships at RPI, NASA-GSFC
and the University of Bern he is now a senior scientist at ETH in Zurich.
\end{IEEEbiography}

\newpage

\appendix

\begin{figure*}[h]
\centering
\includegraphics[width=1.2\columnwidth]{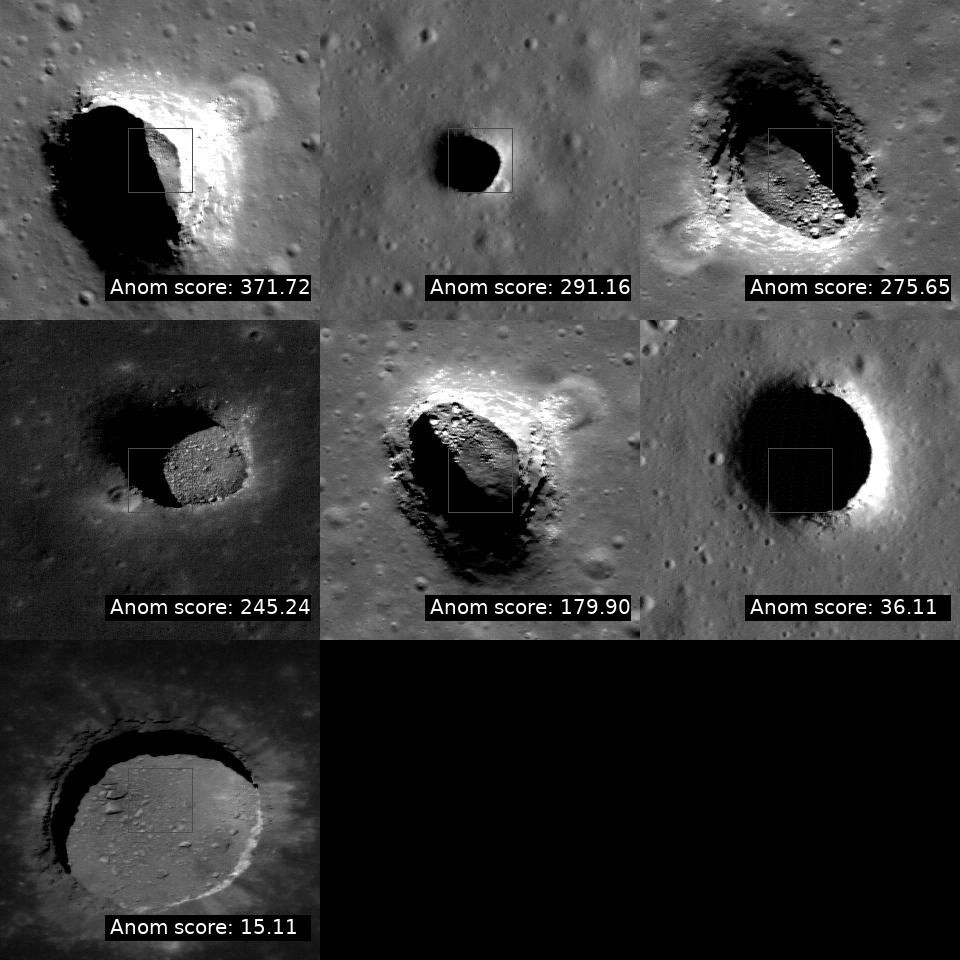}
\includegraphics[width=1.2\columnwidth, trim={0 11cm 0 0},clip]{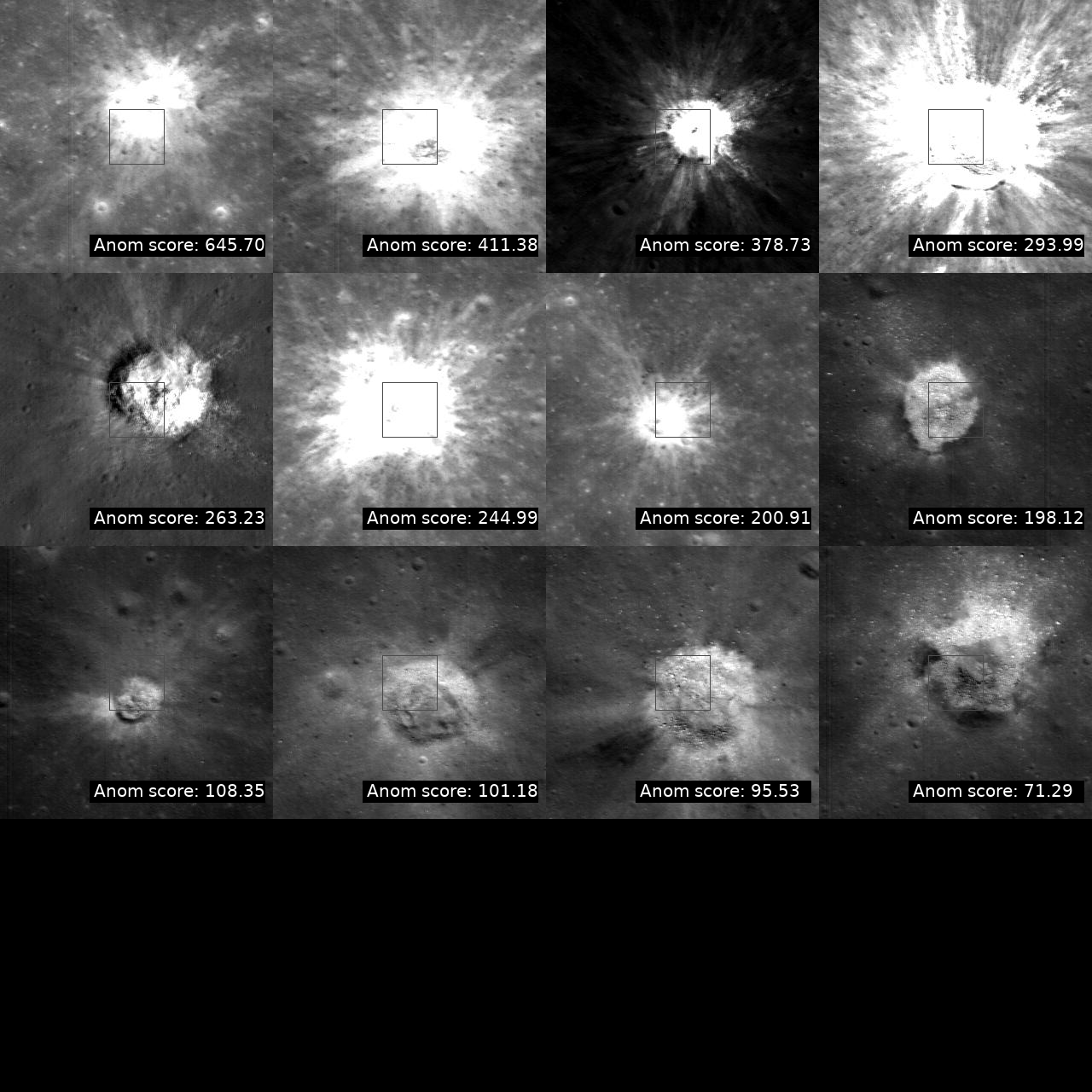}
\caption{All seven and twelve positive labelled patches for pit (top) and crater
(bottom) images, sorted descending by anomaly score in each feature group. We
note that the two pit images with significantly lower anomaly scores than the
rest have the model's input window inside of the (relatively featureless) pit
floor, not intersecting any of the pit's notable and geomorphologically distinct
edges. Raw image credits to NASA/LROC/GSFC/ASU. }
\label{fig:all_positives_crater+pit}
\end{figure*}

\begin{table*}[h]
\centering
\caption{Model versus random baseline average precisions for feature sets, 
    with improvement ratios. Higher AP and improvement factors are better.}
\begin{tabular}{lrrrr}
\toprule
\textbf{Feature} & \textbf{Split} & \textbf{AP} & \textbf{Random AP} & \textbf{Improvement} \\
\textbf{} & \textbf{} & \textbf{} & \textbf{} & \textbf{Factor}     \\
\midrule
\textbf{Craters}        & Test  &  7.54e-3  & 2.30e-5   & 327.40    \\
\textbf{IMP}            & Test  &  2.01e-4  & 1.55e-5   & 12.94     \\
\textbf{Pits}           & Test  &  3.99e-3  & 1.67e-5   & 239.20    \\
\textbf{Rockfalls}      & Test  &  5.70e-5  & 2.27e-5   & 2.51      \\
\textbf{Apollo 12}      & Test  &  5.25e-5  & 1.57e-5   & 3.34      \\
\textbf{Apollo 15}      & Val   &  3.32e-5  & 1.60e-5   & 2.07      \\
\textbf{Apollo 16}      & Test  &  1.06e-4  & 1.55e-5   & 6.80      \\
\textbf{Surveyor 3}     & Test  &  3.46e-5  & 1.57e-5   & 2.21      \\
\bottomrule
\end{tabular}
\vspace{0.25cm}
\label{tab:ap_model_vs_baseline}
\end{table*}

\begin{figure*}[h!t]
\centering
\includegraphics[width=0.9\columnwidth]{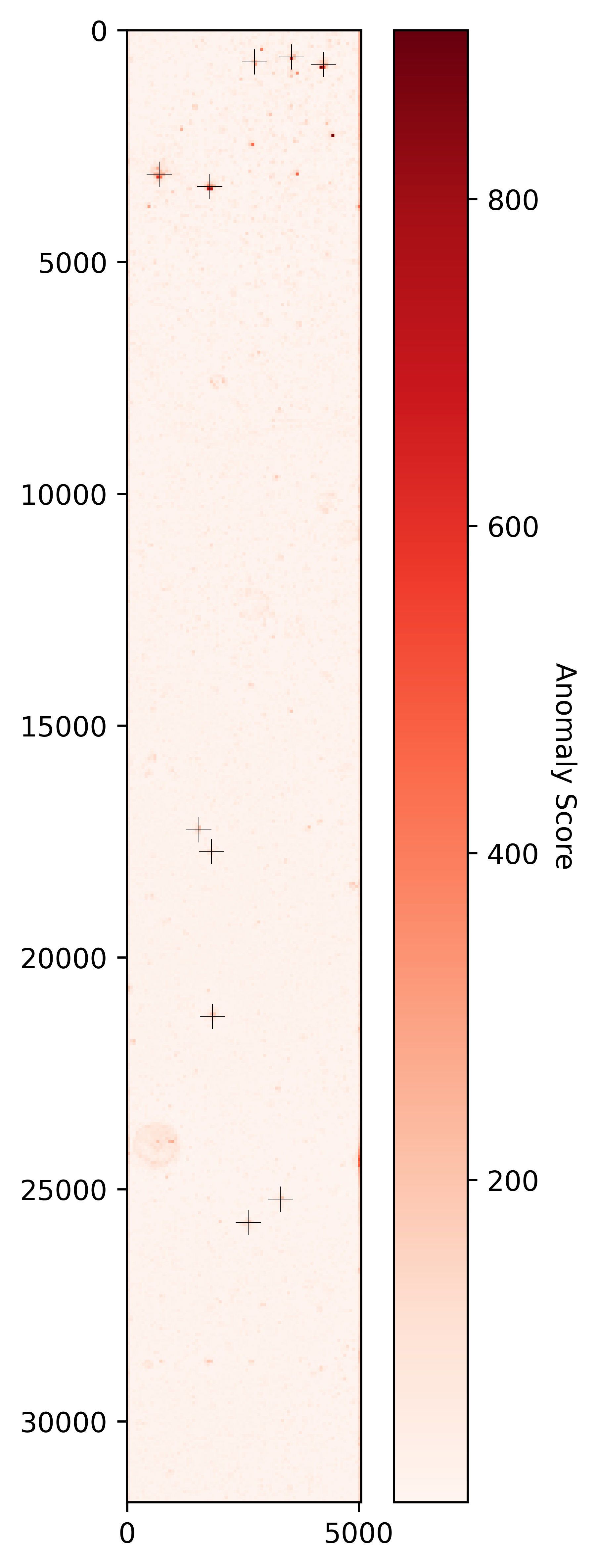}
\includegraphics[width=0.9\columnwidth]{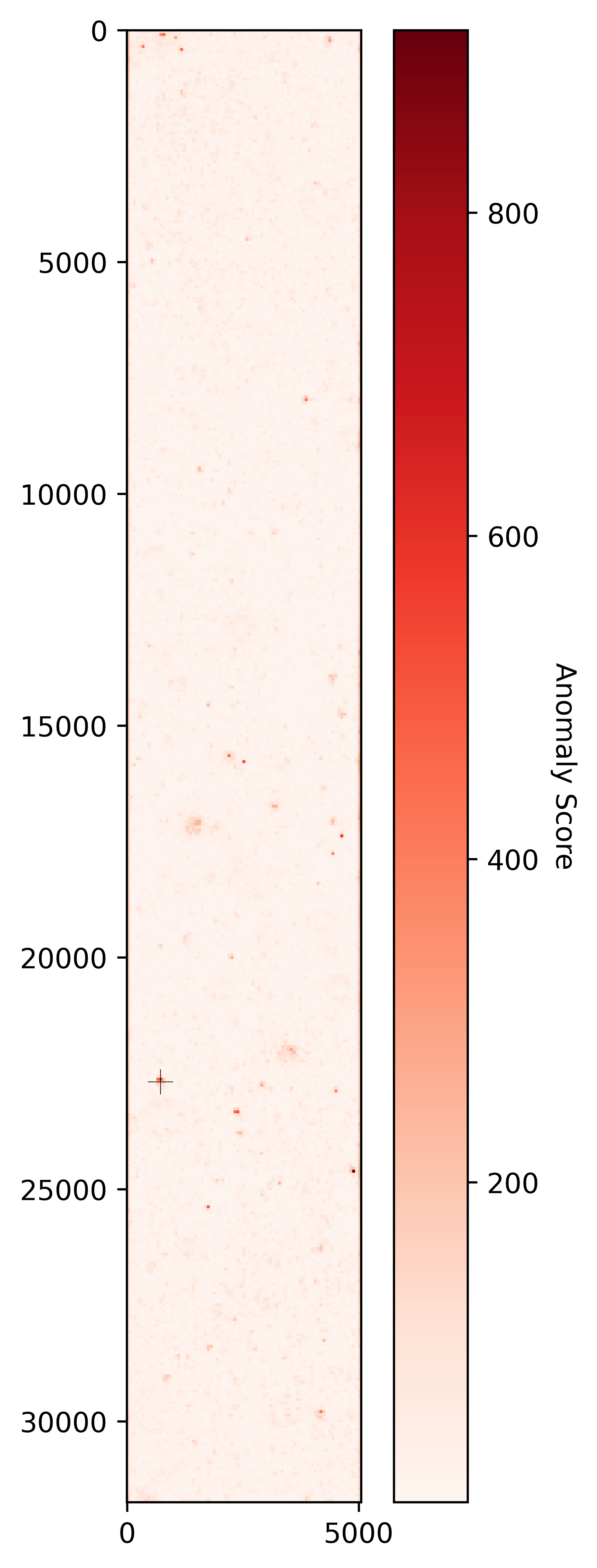}
\caption{Anomaly maps of two LROC images. Black crosses mark positive locations
of features of interest, in this case small fresh impact craters (left) and a
volcanic pit (right). The vast majority of anomalous features without crosses in
the left map are relatively fresh impact craters, but were not marked as `fresh
crater' positives due to their slightly older age (see Discussion section).
 }
\label{fig:anomaly_map}
\end{figure*}

\begin{figure*}[h!t]
\centering
\includegraphics[width=0.95\textwidth]{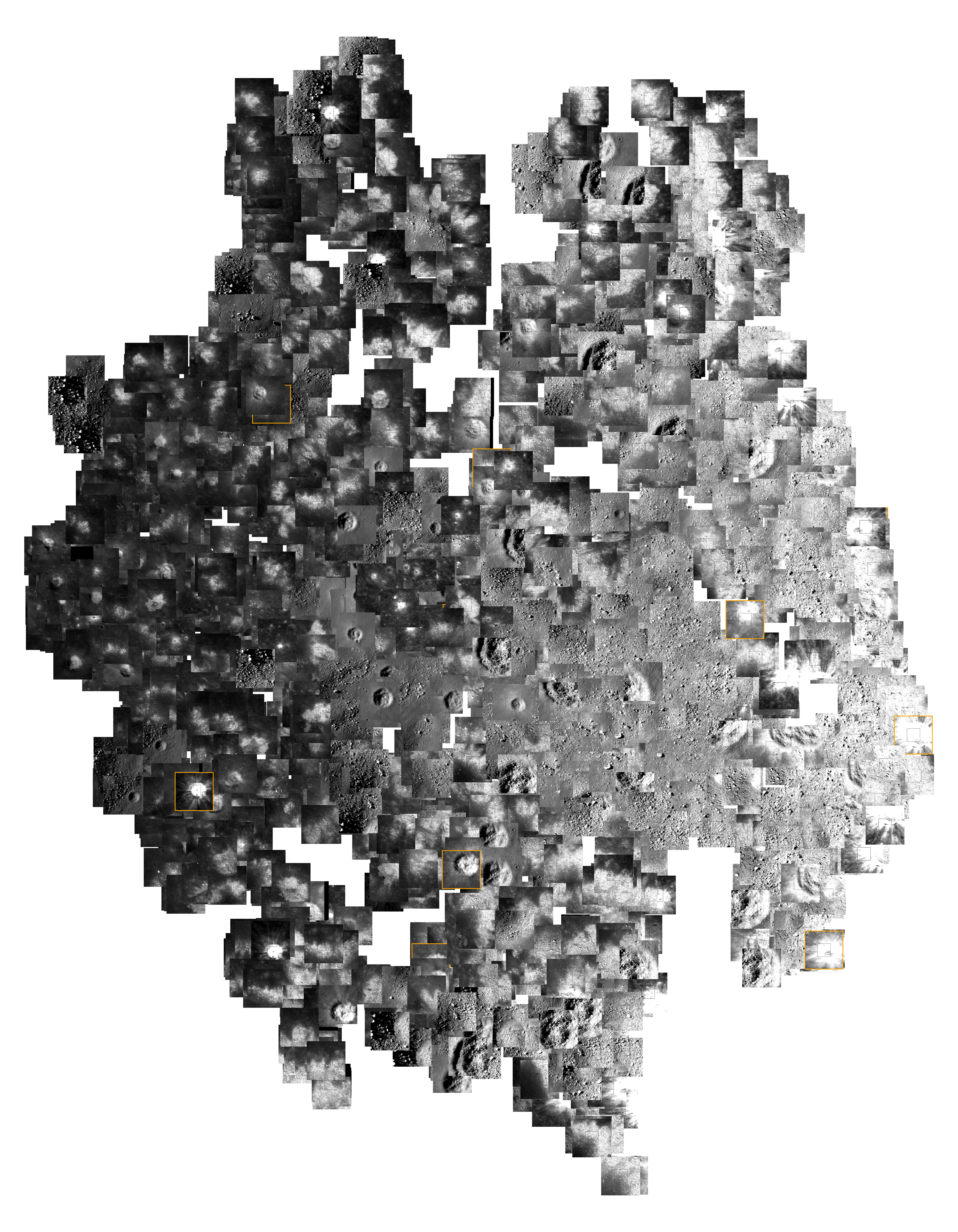}
\caption{t-SNE plot of top 2048 anomalous image patches from the crater test
set. This type of t-SNE plot projects the high-dimensional structure between
images into a two-dimensional representation seen here that can be both easily
visualized while aiming to map images similar in high-dimensional space to
nearby points in this two-dimensional visualization. Note that fresh craters can
be larger than the model's 64 x 64 pixel model window size, so that while each
fresh crater only has one labelled positive sample which is colored here, the
same fresh crater may appear in multiple plotted samples, due to multiple,
different model input windows covering the same fresh crater. Raw image credits
to NASA/LROC/GSFC/ASU. 
}
\label{fig:tsne_crater}
\end{figure*}

\begin{figure*}
    \centering
    \includegraphics[width=0.95\textwidth]{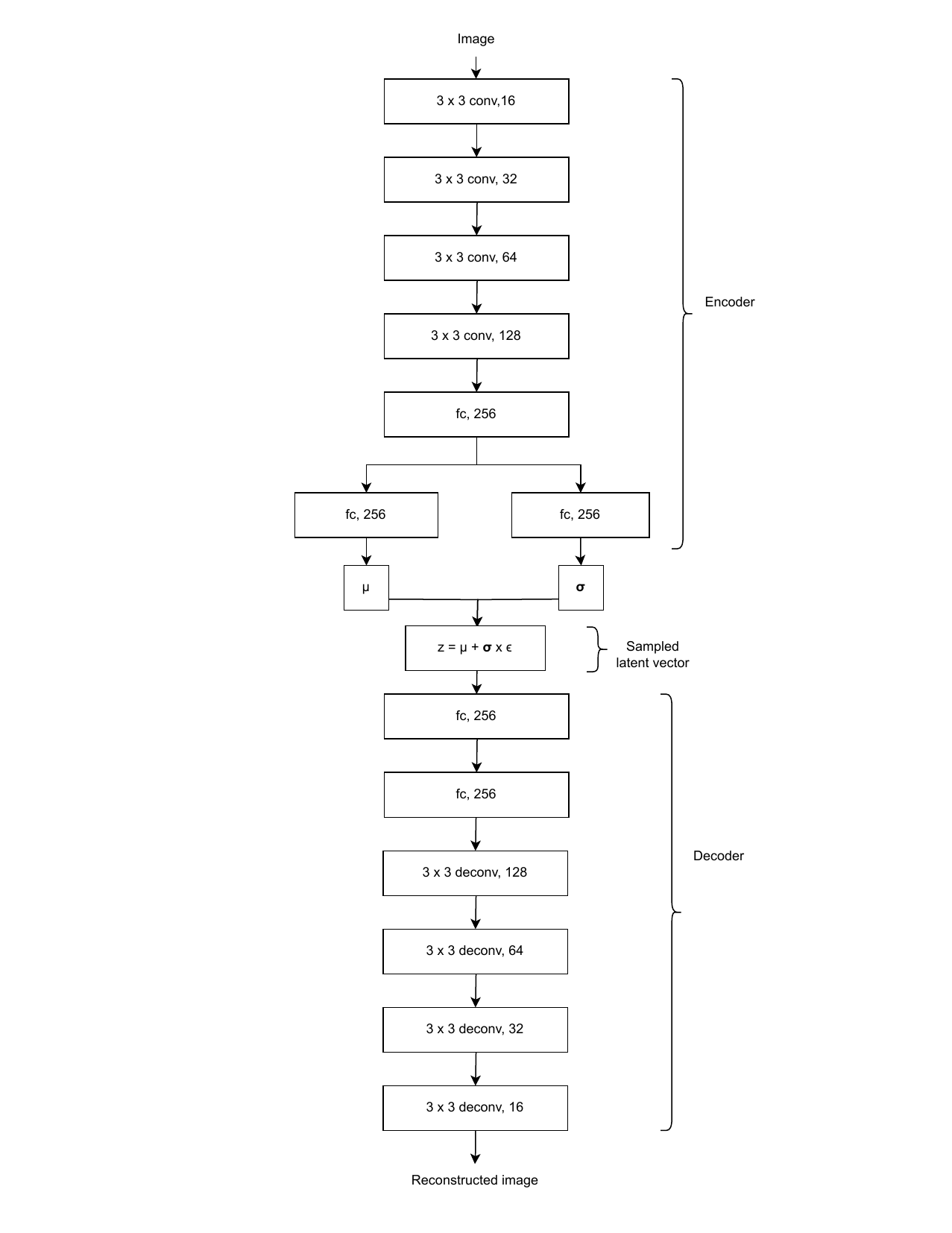}
    \caption{Network diagram. Convolution layers are denoted by ``conv",
    deconvolution layers by ``deconv", and fully connected layers by ``fc".
    Convolution and deconvolution layers are preceded by their window size, e.g.
    ``3 x 3 conv" for a 3-by-3 convolution window. The last number in each box
    denotes either the number of information channels in each convolution or
    deconvolution layer, or the number of nodes in each fully connected layer.
    The $\mu$, $\sigma$, and $z$ latent variables each have dimension 256. Each
    convolution, deconvolution and fully connected layer is followed by a
    batch-norm layer and relu non-linearity, not pictured for clarity.
    }
    \label{fig:network_diagram}
\end{figure*}

\begin{figure*}
    \centering
    \includegraphics{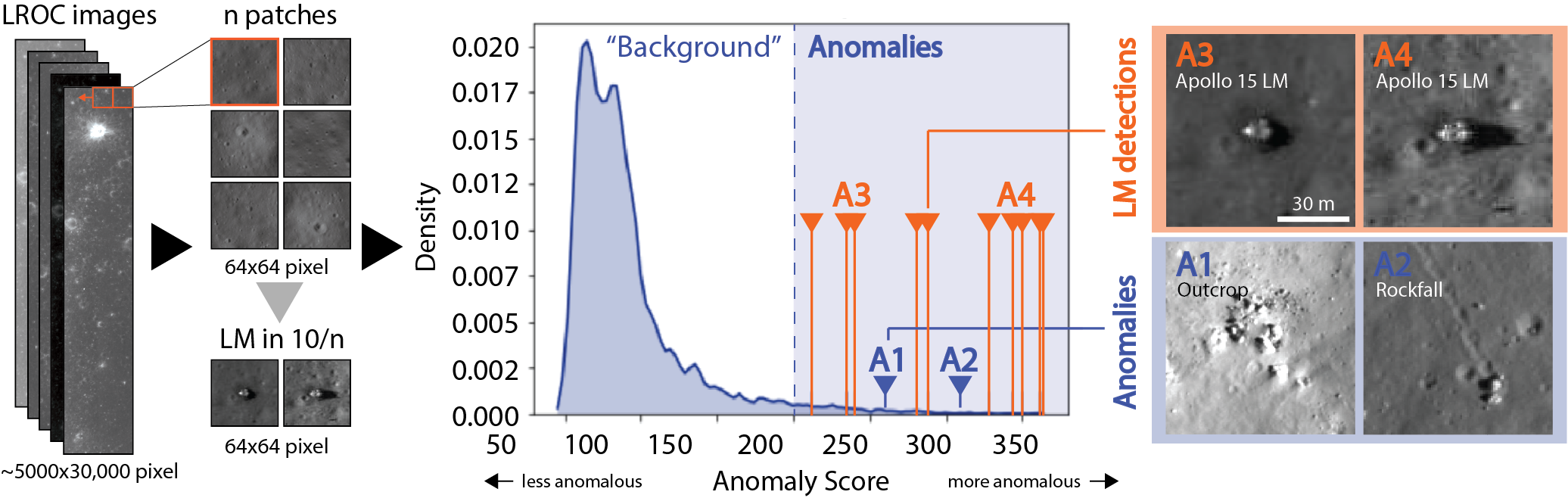}
    \caption{Chart visualizing the overall workflow using all available images
    taken over the Apollo 15 landing site: LROC images are tiled into small
    64x64 pixel patches and processed; the 10 instances of the Lunar Module (LM)
    all cluster on the far right of the anomaly score distribution. Four
    selected anomalies (1 outcrop A1, 1 rockfall A2, and 2 LM A3\&A4) are
    showcased on the right. Note that the boundary between "background" and
    "anomalies" is shown as a hard value for illustrative purposes in this
    diagram. Raw image credits to NASA/LROC/GSFC/ASU. }
    \label{fig:enter-label}
\end{figure*}

\end{document}